\documentclass[manuscript]{aastex}

\slugcomment{Accepted for publication in the Astrophysical Journal.}
\usepackage{amsmath}

\shorttitle{On the Polarization of Inverse Compton Emission}
\shortauthors{Krawczynski}

\def\uga { \raisebox{-0.5ex} {\mbox{$\stackrel{>}{\scriptstyle \sim}$}}}

\def\deg {$^\circ$ }
\def\nerr#1#2 {{+{#1}}-{#2}}
\def\mh#1{{\bf \hat{#1}}}

\newcommand{\plane}[2]{({\bf #1}-{\bf #2})-plane}
\newcommand{\eq}[1]{Equation~(\ref{#1})}
\newcommand{\eqs}[2]{Equations~(\ref{#1}) and (\ref{#2})}
\newcommand{\eqft}[2]{Equations~(\ref{#1}-\ref{#2})}

\begin{document}


\title{The Polarization Properties of Inverse Compton Emission 
and Implications for Blazar Observations with the GEMS X-Ray Polarimeter}


\author{H. Krawczynski\altaffilmark{1}}
\affil{Washington University in St. Louis,
Physics Department and McDonnell Center for the Space Science, 
1 Brookings Drive, CB 1105, St. Louis, MO 63130}
\altaffiltext{1}{Visiting Scientist at the Observatoire de Paris-Meudon 
in Summer 2011, 11, Avenue Marcelin Berthelot, F-92 195 Meudon cedex, France}


\begin{abstract}
NASA's Small Explorer Mission GEMS (Gravity and Extreme Magnetism SMEX),
scheduled for launch in 2014, will have the sensitivity to detect and
measure the linear polarization properties of the 0.5 keV and 
2-10 keV X-ray emission of a considerable number of galactic and extragalactic 
sources. The prospect of sensitive X-ray polarimetry justifies a closer look at 
the polarization properties of the basic emission mechanisms.
In this paper, we present analytical and numerical calculations of the 
linear polarization properties of inverse Compton 
scattered radiation. We describe a generally applicable formalism that can be used
to numerically compute the polarization properties in the Thomson and Klein-Nishina regimes.
We use the code to perform for the first time a detailed comparison of numerical
results and the earlier analytical results derived by Bonometto et al.~(1970) for
scatterings in the Thomson regime. Furthermore, we use the numerical formalism to 
scrutinize the polarization properties of synchrotron self-Compton emission, 
and of inverse Compton radiation emitted in the Klein-Nishina regime. 
We conclude with a discussion of the scientific potential of future GEMS 
observations of blazars. The GEMS mission will be able to confirm the 
synchrotron origin of the low-energy emission component from high frequency 
peaked BL Lac objects. Furthermore, the observations have the potential
to decide between a synchrotron self-Compton and external-Compton origin of 
the high-energy emission component from flat spectrum radio quasars and 
low frequency peaked BL Lac objects.
\end{abstract}





\section{Introduction}
Although the scientific potential of X-ray polarimetry observations has long been 
appreciated \citep[e.g.][]{Rees:75,Ligh:75,Ligh:76,Mesz:88}, 
only a single dedicated X-ray polarimeter mission has been flown so far: the OSO-8 
mission launched in 1975 \citep{Novi:77}. The mission was able to detect polarization 
degrees of a few percent for X-ray bright sources with Crab-like fluxes.
This sensitivity was sufficient to detect the polarization of one cosmic source.
The observations of the Crab nebula revealed evidence for linear polarization at 
2.6~keV and 5.2~keV at a level of $\sim$20\% \citep{Weis:78}.
More recently, the INTEGRAL satellite was used to study the polarization of the Crab 
Nebula and Cyg X-1 \citep{Dean:08,Foro:08,Laur:11}. Even though the observations 
suggest polarization degrees of \uga 40\% in the hard X-ray/$\gamma$-ray regime, 
the systematic uncertainties on these results are large. 
NASA plans to launch the Small Explorer Mission GEMS (Gravity and Extreme Magnetism SMEX) 
in 2014, a dedicated mission for X-ray polarimetry in the 2-10 keV energy range, with the 
sensitivity to detect polarization degrees down to $\sim 1$\% 
even for weak sources with a flux of a few mCrab \citep{Blac:10}. 
A student experiment will extend the polarimetric measurement capabilities 
into sub-keV energy range \citep{Kaar:09}. 
GEMS will be able to measure the polarization properties of dozens of galactic
and extragalactic sources and to establish X-ray polarimetry as an observational 
discipline. The main science drivers of GEMS are the measurement of the
orientation and inclination of the inner accretion disks and the masses and spins of 
Binary Black Hole systems (BBHs), the detection of plasma physics effects 
in neutron star magnetospheres, the study of the locale and properties of the 
particle acceleration regions in pulsars and pulsar wind nebulae, 
and the study of the coronae and jets of Active Galactic Nuclei (AGNs)
\citep[see][]{Lei:97,Weis:06,Bell:10,Kraw:11}.\\[2ex]
The upcoming launch of the GEMS mission motivates a close look at the polarization
properties of the most common X-ray emission processes. Whereas the polarization 
properties of synchrotron emission, Bremsstrahlungs emission, and Thomson scattered emission 
are well studied 
\citep[e.g.][and references therein]{Ginz:65,Dola:67,Lain:80,Bjor:82,Bjor:85,Rybi:86,Bege:93,Rusz:02}, a similar 
in-depth understanding of the polarization properties of inverse Compton emission is still lacking.
In the following, ``Inverse Compton'' denotes the emission from a 
single up-scattering of low-frequency target photons by relativistic electrons.
Prime examples of inverse Compton emission include the high-energy component
of AGNs and possibly $\gamma$-ray emission from Gamma-Ray Bursts. 
This paper focuses on the polarization of inverse Compton emission and
the application of the results to AGN jets.
We will not discuss the polarization of ``Comptonized'' 
emission (e.~g.\ from AGN coronae), which underwent many 
Compton and inverse Compton scatterings before leaving the source. 
Inverse Compton scatterings tend to reduce the polarization degree of emission; 
multiple Compton scatterings do so even more.

The polarization of inverse Compton emission usually arises from averaging 
the results from elemental scattering processes over a range of electron 
energies and propagation directions. Highly relativistic electrons with Lorentz factors $\gamma\gg1$ 
scatter photons into a narrow cone centered on their propagation direction with an opening 
angle of $2/\gamma$. The inverse Compton emission from a relativistic electron plasma thus 
comes from electrons with velocities aligned to within $\sim\gamma^{-1}$ with the line of sight. 
In this paper, we discuss the case that the angular distribution of the 
electrons contributing to the observed emission does not deviate appreciably from 
isotropy in the rest frame of the emitting plasma, called the plasma frame (PF) 
in the following. For large $\gamma$-values most electron populations will satisfy 
this criterium.

The paper of \citet{Bono:70} (called BCS in the following) reports on
the polarization of the inverse Compton emission on the basis of a quantum mechanical scattering 
calculation. The study makes a series of approximations and is limited 
to highly relativistic electrons ($\gamma\gg 1$) and scatterings in 
the Thomson regime $\gamma (\hbar \omega / m_{\rm e} c^2) \ll 1$ where $\omega$ denotes the frequency of the target photons. 
The paper presents equations to compute the polarization degree 
of the inverse Compton emission from isotropic electrons with arbitrary 
energy spectra scattering monoenergetic unidirectional photons. The authors note that the
results imply a vanishing polarization degree for the inverse Compton 
emission from unpolarized target photon beams.
In a follow-up paper, \citet{Bono:73} (referred to as BS in the following) describe a study of 
the polarization of the synchrotron self-Compton (SSC) emission from
electrons in a magnetic field inverse Compton scattering their 
synchrotron emission. The calculation involves the numerical integration of the BCS equations. 
\citet{Nagi:93} confirm the analytical results of BCS and cast the results into the form
of a scattering matrix which operates on the Stokes vector of the target beam.
The authors evaluate the scattering matrix for power law and Maxwellian electron 
distributions, and \citet{Pout:94} uses these results to discuss some aspects 
of the polarization of SSC emission.

The shortcoming of all the analytical calculations is that it is not 
clear at which Lorentz factors the results start to be valid, and how accurate the predicted 
polarization degrees are in any specific case. Various authors have thus developed
codes to numerically evaluate the polarization degrees. 
\citet{Bege:87} describe a formalism to numerically compute the polarization 
of inverse Compton scattered {\em unpolarized} target beams in the Thomson regime. 
They use this formalism to study the polarization of electron beams (``AGN jets'') 
with finite opening angles and with different internal structures (e.~g.\ filled and hollow cones). 
Most of their discussion focuses on situations in which the observer is either 
not looking directly into a modestly relativistic electron beam, or that 
the intensity of the electron beam varies strongly over angular scales of  $\gamma^{-1}$. 

\citet{Celo:94} (called CM in the following) extend the formalism of 
\citet{Bege:87} to treat partially polarized beams and present numerical 
results concerning the polarization of SSC emission from power law electron populations. 
They find that SSC emission is highly polarized - although somewhat less
than the synchrotron emission. The polarization direction of the SSC emission aligns 
with that of the synchrotron emission, i.~e.\ the electric field 
vectors of the synchrotron and SSC emission are perpendicular to the projection 
of the magnetic field lines in the PF onto the plane of the sky. 
While the results depend somewhat on the electron Lorentz factors 
in the mildly relativistic regime, they are largely independent 
of $\gamma$ for $\gamma\uga 10$. 
Their calculations also show that unpolarized synchrotron photons 
give rise to SSC emission with vanishing or very small ($<5\%$) 
polarization degrees for all but very low Lorentz factors.
Although the results of CM deviate by $\sim 15\%$ from those 
of BS, the authors do not investigate the origin of this discrepancy.

More recently, \citet{McNa:09} \citep[see][for additional information on the numerical 
algorithm]{McNa:08} report on simulations of the emission from AGNs for specific 
geometries and for various unpolarized (accretion disk emission, cosmic microwave background (CMB)) 
and polarized (SSC emission) target photon fields. Among other cases, they compute the 
polarization of inverse Compton scattered unpolarized CMB emission. They report a 
polarization of $\sim$20\% in contradiction to the predictions of BCS of a 
vanishingly small net-polarization.

In this paper, we re-evaluate the polarization properties of inverse Compton and SSC emission
based on both analytical and numerical calculations. 
We describe for the first time detailed comparisons of analytical and numerical results. 
The comparisons show in which regimes the analytical approximations can be used. 
Our discussion focuses on establishing the general properties of inverse Compton 
scattered emission. In this spirit, we discuss SSC emission as a special case of 
inverse Compton emission. Although we show some results for mildly relativistic 
electrons ($\gamma$-values of 2 and 5), we focus the discussion on the regime 
$\gamma\uga 10$, where the polarization degree and direction have simple dependencies 
on the relative orientation of the electron momentum, target photon momentum, 
and target photon polarization vectors. 
We use the numerical simulations to investigate the discrepancies between the analytical 
and semi-analytical results of BCS and BS on the one hand, and the numerical 
results of CM and \citet{McNa:09} on the other hand. Furthermore, we present 
for the first time calculations of the polarization degree of inverse Compton 
radiation emitted by isotropic electrons in the Klein-Nishina regime.  

In Section~\ref{S:calculations} we describe a generally applicable formalism that
can be used to simulate inverse Compton emission from arbitrary electron and 
target photon distributions. 
The formalism can be used for arbitrary electron Lorentz factors and 
for scatterings in the Thomson and in the Klein-Nishina regimes, and 
permits the simulation of SSC processes. 
In Sections~\ref{S:analytical} and \ref{S:integrationKernels} we review the results 
from BCS and BS for the Thomson regime, and use them to derive analytical expressions 
for the intensity and polarization of the inverse Compton emission from  
monoenergetic electrons incident on monoenergetic unidirectional target photons. 
In Section~\ref{S:thomson} we perform detailed comparisons of the analytical and numerical
results. The discussion covers the emission from monoenergetic electron and photon beams, 
from isotropic power law electrons scattering monoenergetic unidirectional photons, 
and from isotropic power law electrons scattering power law synchrotron emission. 
We scrutinize the polarization properties of inverse Compton scattered
unpolarized CMB photons in Section \ref{S:CMB}, and study the polarization
properties of inverse Compton radiation emitted in the Klein-Nishina 
regime in Section~\ref{S:kn}. 
Section~\ref{S:discussion} ends the paper with a summary 
and a discussion of the scientific potential of AGN observations 
with GEMS.
\section{Monte Carlo simulations of inverse Compton processes 
in the Thomson and Klein-Nishina regimes}
\label{S:calculations}
\subsection{General approach, Stokes vectors, and useful functions}
\begin{figure}
\resizebox{10cm}{!}{\plotone{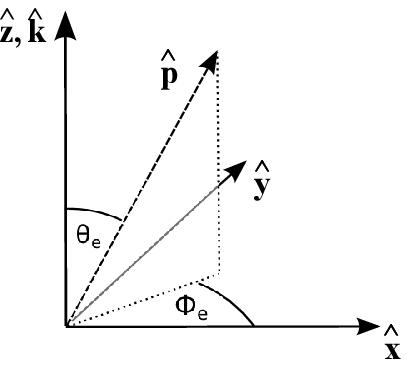}}
\caption{\label{F:pf} Sketch of the simulation set-up in the plasma frame.
An unidirectional beam of electrons propagating along the direction $\mh{p}$ with polar coordinates 
$(\theta_{\rm e},\phi_{\rm e})$ scatters an unidirectional 
photon beam propagating along the $\mh{k}=\mh{z}$.}
\end{figure}

\begin{figure}
\resizebox{10cm}{!}{\plotone{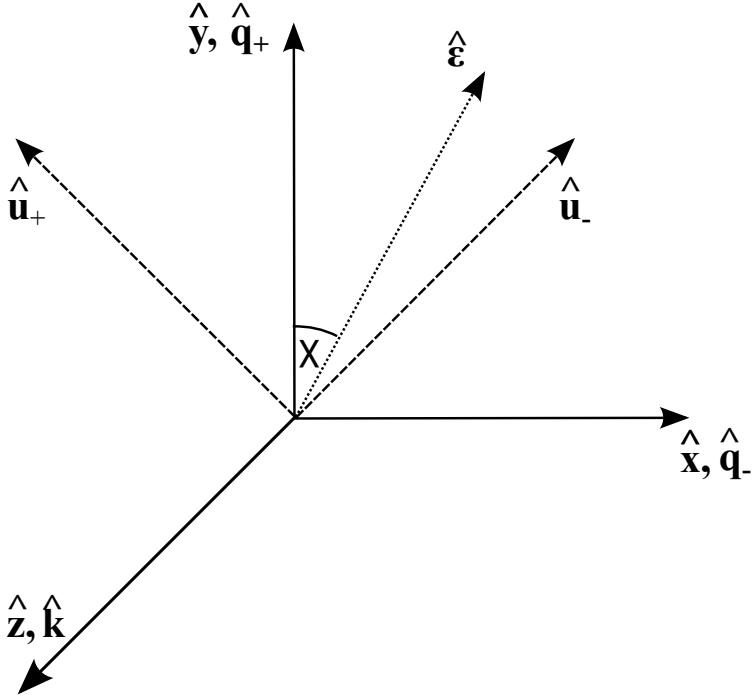}}
\caption{\label{F:stokes} We define the polarization direction
of a photon propagating along $\mh{k}$ with reference to a right handed 
coordinate system $C=\{\mh{x},\mh{y},\mh{z}\}$ with $\mh{z}=\mh{k}$.
The electric field vector of photons with a Stokes parameter 
$q/i=$1 is aligned with the $y$-axis, that of photons
with $q/i=$-1 is aligned with $x$-axis.
Looking along -$\mh{k}$, the directions with $u/i$=1 and $u/i$=-1 are 
rotated counterclockwise by $\pi/4$ relative to the $q/i=$1 and $q/i=$-1 directions,
respectively. The polarization angle $\chi$ is measured clockwise from the 
direction with $q/i=$1.}
\end{figure}

The numerical calculation uses a Monte Carlo approach that avoids some of the 
problems of earlier calculations based on integration over the entire phase space.
Celotti \& Matt (1994) use the latter approach and describe its difficulties:
``For high values of the Lorentz factor $\gamma$, the functions in the integrals 
are strongly peaked in the electron direction. This requires the adoption of a 
carefully sampled angular grid over which to perform the angular grid...''.
In contrast, the Monte Carlo approach generates events with a relative frequency 
matching their natural occurrence and thus makes good use of the computational
resources, even in the extremely relativistic regime.
The simulation of inverse Compton processes resembles in some aspects the simulation of 
Compton scatterings \citep[e.g.][]{Matt:96,Depa:03} with the added complication 
of a Dopper boost from the PF to the electron rest frame (EF) and back.
The calculation begins and ends in the PF. We use a right-handed set
of unit vectors  $C\equiv\{{\bf \hat{x},\,\hat{y},\,\hat{z}} \}$
to define a reference coordinate system in the PF.
Initially, we consider monoenergetic unidirectional electrons with an energy-momentum four-vector 
$p=(\gamma m_{\rm e} c, {\bf p})$ scattering a monoenergetic unidirectional beam
of photons with the four-vector $k=(\omega/c,{\bf k})$ with $\mh{k}=\mh{z}$ (see Figure~\ref{F:pf}).\\[2ex]
%
%
We use unit-less Stokes vectors ${\bf s}=(i,q,u)$ to track information about
the statistical weight ($i$) and the linear polarization properties (via $q/i$ and $u/i$) 
of the simulated photons. Conceptually, we adopt the quantum mechanical treatment of the
Stokes parameters described by \citet{McMa:61}. Accordingly, the Stokes parameters are
expectation values of measurement results on statistical ensembles of incoherent photons. 
The additive property of the Stokes parameters makes them the tool of choice for the 
numerical treatment, as the results from many scattering processes can be combined 
by adding the Stokes vectors. 
\\[2ex]
In the following, we will define Stokes vectors ${\bf s}_{\rm a}$ relative to
various coordinate systems $C_{\rm a}\,\equiv\{{\bf \hat{x}_{\rm a}},{\bf \hat{y}_{\rm a}},
{\bf \hat{z}}_{\rm a}\}$ with the $z$-axes always aligned with the photon propagation 
direction. 
When defining new coordinate systems below, we will only give the $y$ and $z$ unit vectors,
as $\mh{x}_{\rm a}$ can always be computed from $\mh{x}_{\rm a}=\mh{y}_{\rm a}\times\mh{z}_{\rm a}$.
We use the convention that a Stokes vector ${\bf s}_{\rm a}=(1,1,0)$ 
refers to a 100\% linearly polarized photon beam with the electric field 
vector ${\hat{\boldsymbol\varepsilon}}$ parallel to 
${\bf \hat{q}_+}\,\equiv{\bf \hat{y}_{\rm a}}$.
The Stokes vector ${\bf s}_{\rm a}=(1,-1,0)$ refers to linearly 
polarization along ${\bf \hat{q}_-}\,\equiv{\bf \hat{x}_{\rm a}}$
(see Figure~\ref{F:stokes}).
Correspondingly, ${\bf s}_{\rm a}=(1,0,1)$ and ${\bf s}_{\rm a}=(1,0,-1)$ refer 
to the electric field vector parallel to 
${\bf \hat{u}_+}\,\equiv ({\bf \hat{y}_{\rm a}}-{\bf \hat{x}_{\rm a}})/\sqrt{2}$
and ${\bf \hat{u}_-}\,\equiv ({\bf \hat{y}_{\rm a}}+{\bf \hat{x}_{\rm a}})/\sqrt{2}$, 
respectively. 
Following the notation of \citet{McMa:61}, the rotation of the Stokes vector 
${\bf s_0}=(1,1,0)$ by an angle $\chi$ clockwise looking along
${\bf -\hat{k}}$ gives
\begin{equation}
{\bf s}_1\,=\,{\rm \bf M}[\chi]\,
\left(
\begin{array}{c}
1\\
1\\
0
\end{array}     
\right)
\,=\,
\left(
\begin{array}{c}
1\\
\cos{2\chi}\\
-\sin{2\chi}
\end{array}     
\right)
\label{E:mchi}
\end{equation}
where we introduced the matrix
\begin{equation}
{\rm \bf M}[\chi]\,=\,
\left(
\begin{array}{ccc}
1 & 0 & 0 \\
0 & \cos{2\chi} & \sin{2\chi} \\
0 & -\sin{2\chi} & \cos{2\chi} \\
\end{array}     \right). 
\end{equation}
\eq{E:mchi} shows that the angle $\chi$ between the 
${\bf \hat{q}_+}$-direction and the polarization vector is given 
by the expression
\begin{equation}
\chi\,=\,\arctan{\frac{-u}{q}}
\label{E:direction}
\end{equation}
where some care has to be taken to select the right branches 
of the arctan-function to avoid jumps of the inferred $\chi$-values.
The equation for the polarization degree $\Pi$ reads
\begin{equation}
\Pi\,=\,\frac{\sqrt{q^2+u^2}}{i}.
\label{E:degree}
\end{equation}

From the above definitions, it follows that a Stokes vector ${\bf s}_0$ 
transforms as 
\begin{equation}
{\bf s}_1\,=\,{\rm \bf M}[\chi]\,\,{\bf s}_0
\end{equation}
for a coordinate transformation from $C_0$ to $C_1$ resulting from a 
counterclockwise rotation around the $z$-axis of $C_0$ (looking along -$\mh{z_0}$).\\[2ex]
For convenience, we introduce three functions. 
The first function operates on a vector and gives the normalized version of its argument
\begin{equation}
\mh{n}[{\bf a}]\,=\,\frac{\bf a}{\sqrt{\bf a \cdot a}};
\end{equation}
the second gives the component of one vector perpendicular 
to a second vector normalized to unit length
\begin{equation}
\mh{P}[{\bf }a,{\bf b}]\,=\mh{n}[{\bf a}-({\bf a}\cdot\mh{b}) \, \mh{b}];
\end{equation}
the third function gives the angle between the 
$y$-axes of two coordinate systems which share a common $z$-axis:
\begin{equation}
\Delta[C_{\rm a},C_{\rm b}]
\,=\,
-
{\rm sign}(\mh{x}_{\rm a}\cdot\mh{y}_{\rm b})\,
\arccos{(\mh{y}_{\rm a}\cdot\mh{y}_{\rm b})}.
\end{equation}
A positive value indicates that $C_{\rm b}$ results 
from a counterclockwise rotation around $\mh{z}_{\rm a}$.
\subsection{Simulation of one inverse Compton process}
\begin{figure}
\resizebox{10cm}{!}{\plotone{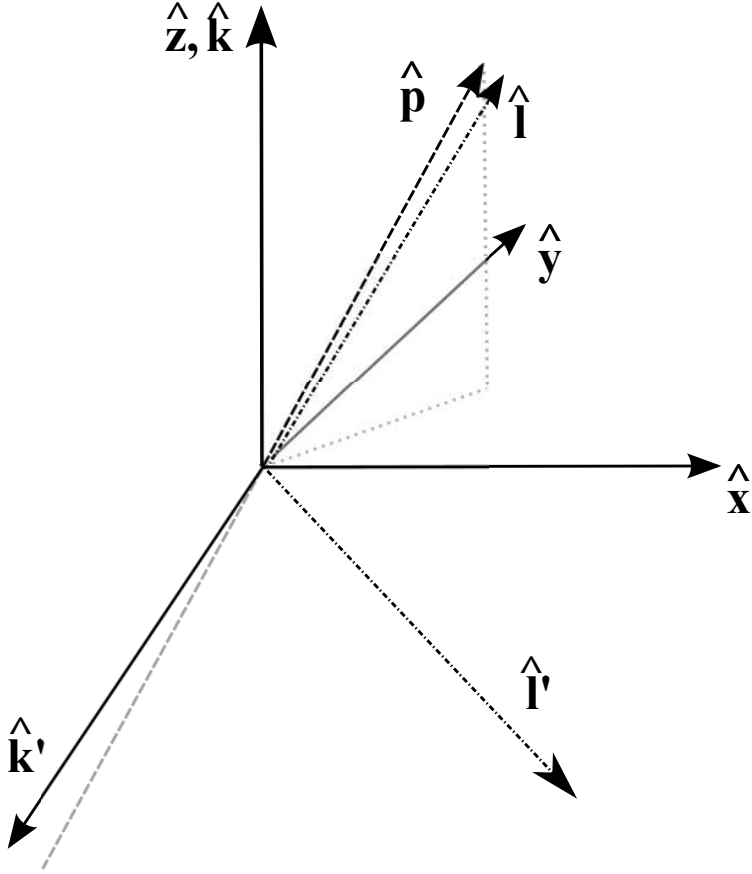}}
\caption{ \label{F:scattering} The Compton scattering process as seen in the 
plasma frame (PF) and in the electron (EF) frame. The PF electron and photon directions before scattering 
are $\mh{p}$ and $\mh{k}$, respectively. The EF photon directions of the target
and scattered photons are $\mh{k}'$ and $\mh{l}'$, respectively. 
The direction of the scattered photon in the PF is denoted with $\mh{l}$. 
}
\end{figure}

%
%
For each photon, an inverse Compton scattering process is simulated with the following 
procedure. See Figure~\ref{F:scattering} for sketches of the scattering 
process in the plasma and electron frames. We dash all quantities in the EF.
\\[1.1ex]
(1) A random number generator is used to draw the direction ${\bf \hat{p}}$ 
($\theta_{\rm e},\phi_{\rm e})$ of the scattering electron
in the PF. We track the likelihood of an electron-photon interaction by 
multiplying the Stokes vector of the photon by the factor (1-$\beta \mu_{\rm kp}$).
Here, $\beta$ is the velocity of the electron in units of the speed of light 
and $\mu_{\rm kp}=\mh{k}\cdot\mh{p}$ is the cosine of the angle between the 
propagation directions of the target photon and the electron.
 \\[1.1ex]
(2) In the second step, we Lorentz transform the photon's wave vector and Stokes vector
into the EF using
\begin{equation}
\mu_{\rm k'p'}'\,=\,\frac{\mu_{\rm kp}-\beta}{1-\mu_{\rm kp}\beta}
\label{E:mu1}
\end{equation} 
and  
\begin{equation}
\mh{k}'\,=\,\mu_{\rm k'p'}' \mh{p}+\sqrt{1-\mu_{\rm k'p'}'{}^2} \,\mh{P}[\mh{k},\mh{p}].
\end{equation}
The photon frequency transforms as
\begin{equation}
\omega'\,=\,\gamma (1-\beta \mu_{\rm kp}) \omega
\label{E:w1}
\end{equation}
with $\gamma=\sqrt{1/(1-\beta^2)}$ being the Lorentz factor of the electron. 

De Young showed that Lorentz boosts do not change 
the linear and circular polarization properties of a statistical ensemble of 
photons including the linear and circular polarization degrees
\citep{DeYo:66}. As we use a Stokes vector to track the statistical weight of 
a scattering process and information about the polarization direction, but not
to track a beam intensity (radiation power per solid angle, area, and time),
the transformed Stokes parameter $i$ is not affected by the Lorentz boosts.
We transform the Stokes vector into the coordinate system
$C_{\rm kp}\,=\{\mh{n}[\mh{k}\times \mh{p}],\mh{k}\}$
with the $y$-axis perpendicular to the \plane{k}{p}:
\begin{equation}
{\bf s}_{\rm kp}={\rm \bf M}[\phi_{\rm e}]\,\, {\bf s}.
\end{equation}
The Stokes vector ${\bf s}_{\rm kp}$ in the PF is identical 
to the Stokes vector ${\bf s}_{\rm k'p}$ defined in the EF 
with regards to the coordinate system
$C_{\rm k'p}\,=\{\mh{n}[\mh{k}'\times \mh{p}],\mh{k'}\}$:
\begin{equation}
{\bf s}'_{\rm k'p}\,=\,{\bf s}_{\rm kp}.
\end{equation}
This important result follows from the Lorentz invariance of the polarization degree and 
from symmetry considerations alone. The latter imply that a beam polarized in the 
({\bf k}-{\bf p})-plane in the PF will be polarized in the same plane in the EF, 
because there is no preferred polarization direction perpendicular to the plane.
Thus, the invariance of the polarization degree implies that a beam with 
$q=-x$ and $u=0$ in the PF will have $q=-x$ and $u=0$ in the EF, i.e.\
that $q$ is Lorentz invariant. The Lorentz invariance of $\Pi$ and $q$ implies 
the Lorentz invariance of $u$. 
Equations (6a-c) of \citep{Bege:87} imply the same result, but were given without
justification.\\[1.1ex]
(3) The third step implements the Compton scattering process in the EF.
We denote the wave vector of the scattered photon with $l'$, and use a random number
generator to draw a random direction for the spatial part of $l'$
assuming a constant probability per solid angle. With $\mu_{\rm k'l'}'=$
$\mh{k}'\cdot \mh{l}'$, the scattering angle is $\theta_{\rm s}'=$ $\arccos{\mu_{\rm k'l'}'}$.
We use Compton's scattering formula $\Delta \lambda=\frac{h}{m_{\rm e}\,c}(1-\mu_{\rm k'l'}')$ 
to calculate the frequency of the 
scattered photon:
\begin{equation}
\omega_{\rm s}'\,=\,\frac{1}{1+\epsilon'\,(1-\mu_{\rm k'l'}')}\omega'
\label{E:w2}
\end{equation}
where $\epsilon'=\hbar \omega'/m_{\rm e}c^2$ is the target photon 
energy in the EF in units of the electron rest mass.
We can now use the part of Fano's Compton scattering matrix relevant
for linear polarization to calculate the Stokes vector of the scattered photon
\citep{McMa:61,Fano:49,Fano:57}:
\begin{equation}
{\bf s}_{\rm l'k'}'\,=\,{\rm \bf F}[\theta_{\rm s}',\epsilon',\epsilon_{\rm s}']\,\,
{\rm \bf M}[\Delta[C_{\rm k'p},C_{\rm l'k'}]]\,\, {\bf s}_{\rm k'p}'.
\end{equation}
The matrix {\bf M} transforms 
the Stokes vector ${\bf s}_{\rm k'p}'$ 
into the coordinate frame 
$C_{\rm l'k'}=\{\mh{n}[\mh{l}'\times\mh{k}'],\mh{l}'\}$
with the $y$-axis perpendicular to the \plane{k$'$}{l$'$} -- a prerequisite for using
Fano's matrix.
Fano's matrix {\bf F} describes Thomson scattering in the Thomson and Klein-Nishina
regimes and reads:
\begin{equation}
{\rm \bf F}[\theta_{\rm s}',\epsilon',\epsilon_{\rm s}']\,=\,
\left(\frac{\epsilon_{\rm s}'}{\epsilon'}\right)^2
\left(
\begin{array}{ccc}
1+{\rm cos}^2\theta_{\rm s}'+(\epsilon'-\epsilon_{\rm s}')(1-\cos{\theta_{\rm s}'}) & {\rm sin}^2\theta_{\rm s}'&0\\
{\rm sin}^2\theta_{\rm s}' & 1+{\rm cos}^2\theta_{\rm s}'&0 \\
0&0&2\cos{\theta_{\rm s}'} 
\end{array}
\right)
\label{E:fano}
\end{equation}
with $\epsilon_{\rm s}'=\hbar \omega_{\rm s}'/m_{\rm e}c^2$ being the 
energy of the scattered photon in the EF in units of the electron rest mass. 
We did not include here the numerical factor $r_0{}^2/2$ (with $r_0$ being the classical
electron radius) in the definition of {\bf F}, as we are not interested in 
absolute fluxes in the following. Note that Fano's matrix accounts for the 
statistical weight of the scattering process by modifying the $i$-component
of the Stokes vector; the resulting Stokes vector is defined relative to $C_{\rm l'k'}$.
\\[1.1ex]
(4) The final step consists in back-transforming the results into the PF.
The cosine of the angle between the scattered photon direction 
and the electron direction $\mu_{\rm l'p'}'=$$\mh{l}'\cdot\mh{p}'$ transforms as
follows into the PF:
\begin{equation}
\mu_{\rm lp}\,=\,\frac{\mu_{\rm l'p'}'+\beta}{1+\mu_{\rm l'p'}'\beta}.
\end{equation}
The scattered photon direction in the PF is
\begin{equation}
\mh{l}\,=\,\mu_{\rm lp} \mh{p}+\sqrt{1-\mu_{\rm lp}{}^2} \,\mh{P}[\mh{l}',\mh{p}].
\end{equation}
The PF frequency of the scattered photon is
\begin{equation}
\omega_{\rm s}\,=\,\gamma(1+\beta\mu_{\rm l'p'}')\omega_{\rm s}'.
\label{E:w3}
\end{equation}

After inferring the Stokes vector of the scattered photon into the coordinate system 
$C_{\rm l'p}=\{\mh{n}[\mh{l}'\times \mh{p}],\mh{l}'\}$
with the $y$-axis perpendicular to the ({\bf l$'$}-{\bf p})-plane:
\begin{equation}
{\bf s}_{\rm l'p}'\,=\,{\rm \bf M}[\Delta[C_{\rm l'k'},C_{\rm l'p}]]\,\, {\bf s}_{\rm l'k'}',
\end{equation}
we Lorentz transform the Stokes vector into the PF relative to 
$C_{\rm lp}=\{\mh{n}[\mh{l}\times \mh{p}],\mh{l}\}$
\begin{equation}
{\bf s}_{\rm lp}\,=\,{\bf s}_{\rm l'p}'.
\end{equation}
Finally, we compute the Stokes vector relative to the coordinate system 
$C_{\rm lk}=\{\mh{P}[\mh{k},\mh{l}],\mh{l}\}$ with the $y$-axis aligned with the 
projection of the target photon's propagation direction in the sky: 
\begin{equation}
{\bf s}_{\rm lk}\,=\,{\rm \bf M}[\Delta[C_{\rm lp},C_{\rm lk}]]\,\,{\bf s}_{\rm lp}.
\end{equation}
\subsection{Simulation of SSC emission}
\label{S:sc}
The simulation of SSC emission from electron and photon powerlaw distributions 
proceeds along similar lines. The calculations assume a uniform magnetic field 
{\bf B} oriented at an angle $\vartheta_{\rm B}$ to the line of sight.
A random photon propagation direction $\mh{k}$ is drawn assuming a constant 
probability per solid angle. Subsequently, we generate a random photon frequency 
between $\omega_1$ and $\omega_2$ from a power law distribution with 
$dN_{\gamma}/dE_{\gamma}\propto E_{\gamma}^{-(\alpha+1)}$. Similarly, an electron
Lorentz factor is drawn from a power law distribution with 
$dN_{\rm e}/d_{\gamma}\propto \gamma^{-p}$.
The initial Stokes vector is set to 
\begin{equation}
{\bf s}_{\rm kB}\,=\,
(1-\beta\mu_{\rm kp}) \,
{\rm sin}^{\alpha+1}{\theta_{\rm B}}\,
\left(
\begin{array}{c}
1\\
\Pi_{\rm S}\\
0
\end{array}
\right)
\end{equation}
with $\theta_{\rm B}$ being the angle between $\mh{k}$ and the magnetic field {\bf B}
and $\mu_{\rm kp}=\mh{k}\cdot\mh{p}$ as above.
The two multiplicative factors account for the relative likelihood of a electron-photon
interaction and for the scaling of the synchrotron emissivity with
$\theta_{\rm B}$ \citep{Ginz:65}. We assume that the emission is linearly polarized along the 
$\mh{q}_+$-direction and the value of the $q$-parameter gives the polarization 
degree of synchrotron emission with a power law index $\alpha$:
\begin{equation}
\Pi_{\rm S}\,=\,\frac{1+\alpha}{\alpha+5/3}.
\end{equation}
As synchrotron emission is polarized perpendicular to the {\bf B}-field,
${\bf s}_{\rm kB}$ is defined with regards to the coordinate system 
$C_{\rm kB}=\{\mh{n}[\mh{k}\times{\bf B}],\mh{k}\}$.
The rest of the procedure is now equivalent to the one described in Steps (2)-(4) above
with the modification that the Stokes vector ${\bf s}_{\rm kp}$ is now given by
\begin{equation}
{\bf s}_{\rm kp}\,=\,{\rm \bf M}[\Delta[C_{\rm kB},C_{\rm kp}]] \,\,{\bf s}_{\rm kB}.
\end{equation}
At the very end of the calculation, we compute the Stokes vector relative to a 
coordinate system $C_{\rm lB}=\{\mh{P}[{\bf B},\mh{l}],\mh{l}\}$
with the $y$-axis along the {\bf B}-field direction in the PF 
projected onto the sky: 
\begin{equation}
{\bf s}_{\rm lB}\,=\,{\rm \bf M}[\Delta[C_{\rm lp},C_{\rm lB}]]\,\,{\bf s}_{\rm lp}.
\end{equation}
\section{Analytical and numerical results}
\label{S:results}
\subsection{Analytical results derived by BCS and BS}
\label{S:analytical}
We continue to use the same naming conventions as in the previous section.
The frequency of an inverse Compton scattered photon can be computed from
Equations~(\ref{E:w1}), (\ref{E:w2}), and (\ref{E:w3}). 
In the Thomson regime $\epsilon'\rightarrow 0$ we get
\begin{equation}
\omega_{\rm s}\,=
\gamma^2 (1-\beta \mu_{\rm kp})(1+\beta \mu'_{\rm l'p'})
\omega.
\end{equation}
This equation gives the following minimum and maximum scattered photon frequencies
as function of $\gamma$, $\omega$, and $\mu_{\rm kp}$:
\begin{equation}
\omega_{\rm min}\,=
\frac{1-\beta \mu_{\rm kp}}{2}
\omega
\end{equation}
and
\begin{equation}
\omega_{\rm max}\,=
2\gamma^2 (1-\beta \mu_{\rm kp})
\omega,
\end{equation}
respectively. We will see below, that $\omega_{\rm max}$ sets a useful scale 
to characterize the frequency dependence of the intensity and polarization degree
of inverse Compton emission. 

From the last equation we infer that the minimum Lorentz 
factor $\gamma_{\rm min}$ required to produce photons of frequency $\omega_{\rm s}$ is
for $\gamma\gg 1$ and $\beta\approx 1$:
\begin{equation} 
\gamma_{\rm min}\,=\,
\sqrt{
\frac{1}{2}
\frac{\omega_{\rm s}}{\omega}
\frac{1}{(1-\mu_{\rm kp})}}.
\label{E:gmin}
\end{equation}

BCS and BS derived an expression for the intensity $J_{\mh{\boldsymbol\varepsilon}_{\rm s}}$
(energy emitted per unit time, unit volume and unit frequency interval) of a unidirectional 
monoenergetic photon beam polarized along the vector $\mh{\boldsymbol\varepsilon}$
scattered by an isotropic electron population in the direction $\mh{l}$ into a state 
of linear polarization defined by the polarization vector $\mh{\boldsymbol\varepsilon}_{\rm s}$:
\begin{equation}
J_{\mh{\boldsymbol\varepsilon}_{\rm s}}\,=\,K\,\gamma_{\rm min} \frac{\omega}{\omega_{\rm s}}
 \left\{ \left[ \mh{\boldsymbol\varepsilon}\cdot\mh{\boldsymbol\varepsilon}_{\rm s}+
\frac{(\mh{k}\cdot\mh{\boldsymbol\varepsilon}_{\rm s})(\mh{l}\cdot\mh{\boldsymbol\varepsilon})}
{1-(\mh{k}\cdot\mh{l})} \right]^2\,
(\Sigma_1+\Sigma_2)+\Sigma_2\right\}
\label{E:sd}
\end{equation}
with $K\equiv$ $\frac{1}{2}h\,c\,r_0{}^2\,\rm cm^{-3}$
and $\gamma_{\rm min}$ as defined above. 
The factors $\Sigma_1$ and $\Sigma_2$ 
depend on the electron energy spectrum:
\begin{equation}
\Sigma_1\,=\int_{0}^{1}\,m(\gamma)\,\left(x^2-\frac{1}{x^2}+2\right) dx
\label{E:s1}
\end{equation}
\begin{equation}
\Sigma_2\,=\int_{0}^{1}\,m(\gamma)\,\frac{(1-x)^2}{x^2} dx.
\label{E:s2}
\end{equation}
The function $m$ depends on $dn_{\rm e}/d\gamma$, the differential electron spectrum 
per unit volume, as: 
\begin{equation}
m(\gamma)=\frac{dn_{\rm e}}{d\gamma}\gamma^{-2},
\end{equation}
and $x$ is defined as
\begin{equation}
x=\frac{\gamma_{\rm min}}{\gamma}.
\end{equation}
Summing Equation (\ref{E:sd}) over two orthogonal polarization directions gives the total
intensity of the inverse Compton scattered radiation:
\begin{equation}
J_{\mh{\boldsymbol\varepsilon}_{\rm s}}+
J_{\mh{\boldsymbol\varepsilon}_{\rm s,\perp}} \,=\,
K\,\gamma_{\rm min}\,\frac{\omega}{\omega_{\rm s}}
 (\Sigma_1+3\,\Sigma_2).
\label{E:sd2}
\end{equation}

BCS evaluated the linear polarization fraction for three exemplary cases with 
different initial polarization directions relative to the scattering plane 
($\mh{\boldsymbol\varepsilon}_{\rm s}$ parallel, at 45$^{\circ}$, and perpendicular
to the scattering plane).
The main results of the calculations are:\\[1.1ex]
(i) The intensity of the inverse Compton radiation is independent of the
polarization degree and direction of the target photon beam, and does not depend on 
the scattering direction.\\[1.1ex]
(ii) The linear polarization degree of the inverse Compton radiation depends 
on the degree but not the direction of the polarization of 
the target photons, and does not depend on the scattering direction.
For a 100\% polarized target beam, the linear polarization degree of the 
inverse Compton scattered radiation is
\begin{equation}
\Pi_{\rm BCS}\,=\,\frac{\Sigma_1+\Sigma_2}{\Sigma_1+3\,\Sigma_2}.
\label{E:bcs}
\end{equation}
Note that although $\Sigma_1<0$ is possible, the combinations 
$\Sigma_1+\Sigma_2$ and $\Sigma_1+3\Sigma_2$ are 
equal or larger than zero.
BCS evaluated $\Pi_{\rm BCS}$ analytically for powerlaw electron 
distributions and numerically for the case of SSC 
emission.
\\[1.1ex]
(iii) In contrast to Thomson scattering by stationary electrons,
Inverse Compton scattering by non-thermal electron plasmas does not 
create polarization, and an initially unpolarized beam stays 
unpolarized after scattering. This result implies that if  
inverse Compton scattering reduces the polarization degree 
of a 100\% polarized target beam to $\Pi_1$, it will reduce
the polarization degree of a target beam with a polarization degree of
$\Pi_2$ to a net-polarization of $\Pi_1\,\Pi_2$.
\subsection{Additional analytical results}
\label{S:integrationKernels}
\eq{E:sd} allows us to derive a closed expression for the polarization direction 
of the inverse Compton scattered radiation.
The latter is given by the unit vector $\mh{\boldsymbol\varepsilon}_{\rm s}$ 
which maximizes the term in square brackets of \eq{E:sd} with the constraining 
condition that $\mh{\boldsymbol\varepsilon}_{\rm s}\perp \mh{l}$.
If we pull out the vector $\mh{\boldsymbol\varepsilon}_{\rm s}$ from the 
term in \eq{E:sd} in square brackets, we see that the condition 
is equivalent to minimizing the angular 
distance between $\mh{\boldsymbol\varepsilon}_{\rm s}$ and the vector 
\begin{equation}
{\bf v}\,=\, 
\mh{\boldsymbol\varepsilon}+
\frac{\mh{l}\cdot\mh{\boldsymbol\varepsilon}}
{1-(\mh{k}\cdot\mh{l})}\,\mh{k}.
\label{E:pd1}
\end{equation}
The angular distance is minimized for the projection of ${\bf v}$ 
onto the plane perpendicular to $\mh{l}$. The polarization vector is 
thus given by the expression
\begin{equation}
\mh{\boldsymbol\varepsilon}_{\rm s}\,=\,\mh{P}[{\bf v},\mh{l}].
\label{E:pd2}
\end{equation}
Using this expression, it can be shown in a somewhat tedious calculation 
that the polarization angle
$\chi=\arccos{(
\mh{\boldsymbol\varepsilon}_{\rm s}\cdot \mh{P}[\mh{k},\mh{l}])}$
equals the angle between the initial polarization direction and the 
projection of $\mh{l}$ onto the plane perpendicular to $\mh{k}$:
\begin{equation}
\chi\,=\,\arccos{
(\mh{\boldsymbol\varepsilon}\cdot \mh{P}[\mh{l},\mh{k}])}.
\label{E:pd4}
\end{equation}
While \eq{E:pd2} gives the polarization direction, 
\eq{E:pd4} gives only the absolute value of $\chi$ but not the sign of $\chi$.
With regards to Figure \ref{F:pf}, the polarization angle $\chi$ is $\pi/2-\phi_{\rm e}$.
Note that the polarization direction does not depend on the observed frequency.

The BCS equations allow us to derive the frequency dependence of the intensity 
and polarization degree of the inverse Compton emission from monoenergetic electrons. 
The differential energy spectrum of a monoenergetic electron 
beam normalized to one electron per cm$^3$ is 
\begin{equation}
\frac{dn_{\rm e}}{d\gamma}\,=\, \delta(\gamma-\gamma_0)\,\rm cm^{-3}.
\end{equation}
The integrands in \eqs{E:s1}{E:s2} only contribute for
\begin{equation}
x_0\,\equiv\,\frac{\gamma_{\rm min}}{\gamma_0}.
\label{E:x0}
\end{equation}
We get after some algebra
\begin{equation}
\Sigma_1\,=\,\frac{2-\frac{1}{x_0^2}+x_0^2}{\gamma_0{}^3}
\end{equation}
and
\begin{equation}
\Sigma_2\,=\,\frac{(1-x_0{\,}^2)^2}{x_0\,\gamma_0{}^3}.
\end{equation}
Note that both parameters $\Sigma_1$ and $\Sigma_2$ depend through 
$x_0$ (and thus through $\gamma_{\rm min}$) on the observed frequency $\omega_{\rm s}$.
We obtain the total intensity
\begin{equation}
J\,=\,K\,\frac{\omega_{\rm s}}{\omega}\,
\frac{2\left(1-2 x_0{}^2+2 x_0{}^4\right) }{\gamma_0{}^2}
\end{equation}
and the polarization degree
\begin{equation}
\Pi_{\rm BCS}\,=\,
\frac{x_0{}^4}{1-2 x_0{}^2+2 x_0{}^4}.
\end{equation}
We now introduce the dimensionless variable $y$ as the observed 
frequency in units of the maximum frequency $\omega_{\rm max}(\gamma_0,\mu_{\rm kp})$
which electrons of Lorentz factor $\gamma_0$ emit in a certain direction:  
\begin{equation}
y\,=\,\frac{\omega_{\rm s}}{\omega_{\rm max}}.
\end{equation}
The intensity scales with $y$ as
\begin{equation}
J\,\propto\,y\frac{1-2 y+2 y^2 }{\gamma_0{}^2}.
\end{equation}
We infer a differential photon number per unit frequency interval of
\begin{equation}
\rho_{\gamma_0}\,\equiv\,\frac{dN_{\gamma}}{d\omega_{\rm s}}\,=\,
\frac{dN_{\gamma}}{dy}\frac{dy}{d\omega_{\rm s}}\,
\,=\,\frac{3(1-2 y+2 y^2)}{2 \,\omega_{\rm max}}
\label{E:in}
\end{equation}
where we introduced the normalization factor $3/(2\,\omega_{\rm max})$ so that
$\int_{\omega_{\rm min}}^{\omega_{\rm max}} \rho_{\gamma_0} d\omega_{\rm s}=1$
for $\omega_{\rm min} \approx 0$.

The polarization degree as function of $y$ is
\begin{equation}
\Pi_{\gamma_0}\,=\,
\frac{y^2}{1-2 y+2 y^2}.
\label{E:pi}
\end{equation}
Equations (\ref{E:in}) and (\ref{E:pi}) give the intensity and polarization fraction
of the inverse Compton emission from monoenergetic electrons and can be used
as integration kernel to predict the spectral shape and polarization properties of
the inverse Compton emission from more complicated electron distributions.
\subsection{Comparison of numerical and analytical results in the Thomson Regime}
\label{S:thomson}
\begin{figure}
\hspace*{-3.5cm}
\begin{minipage}{8cm}
\resizebox{11cm}{!}{\plotone{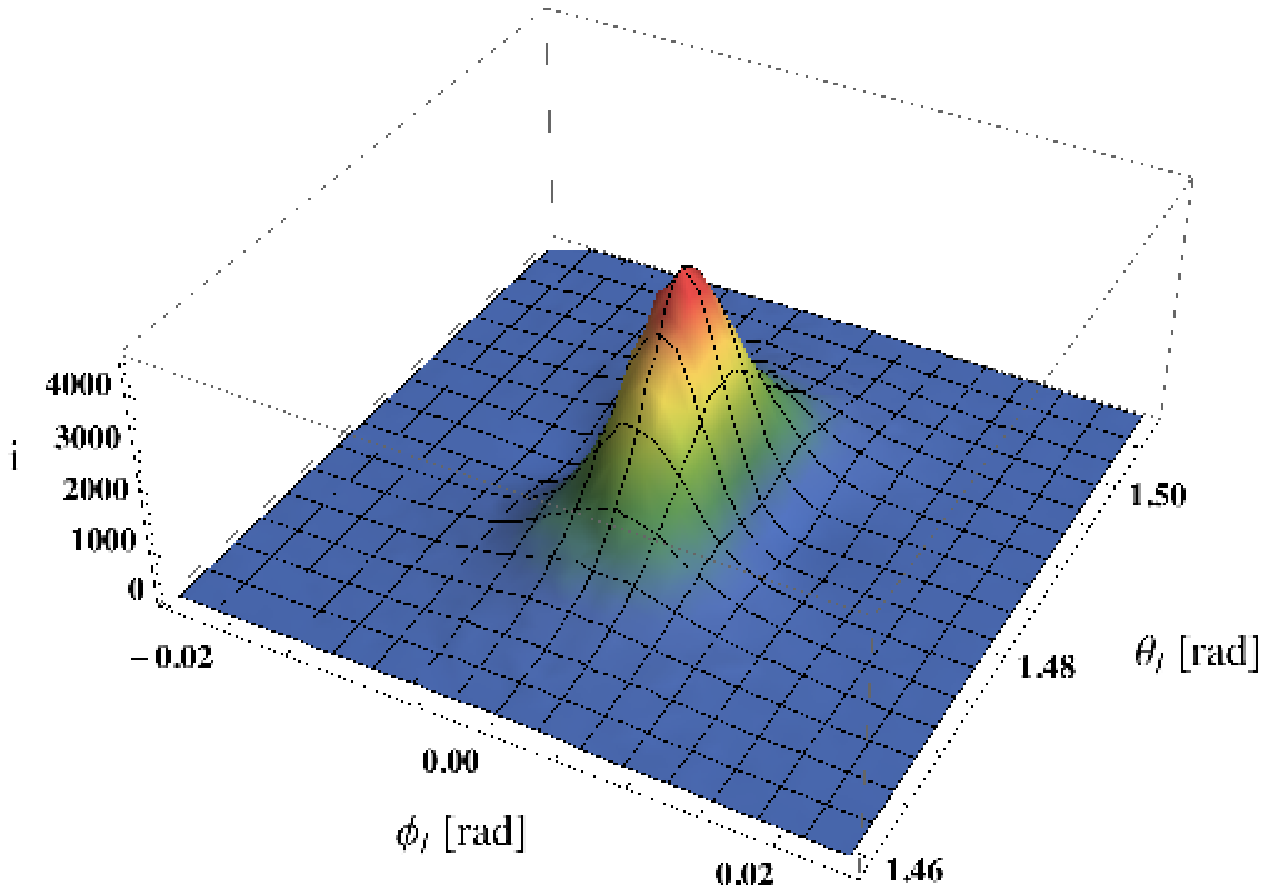}}
\end{minipage}
\begin{minipage}{8cm}
\hspace*{-1cm}
\resizebox{11cm}{!}{\plotone{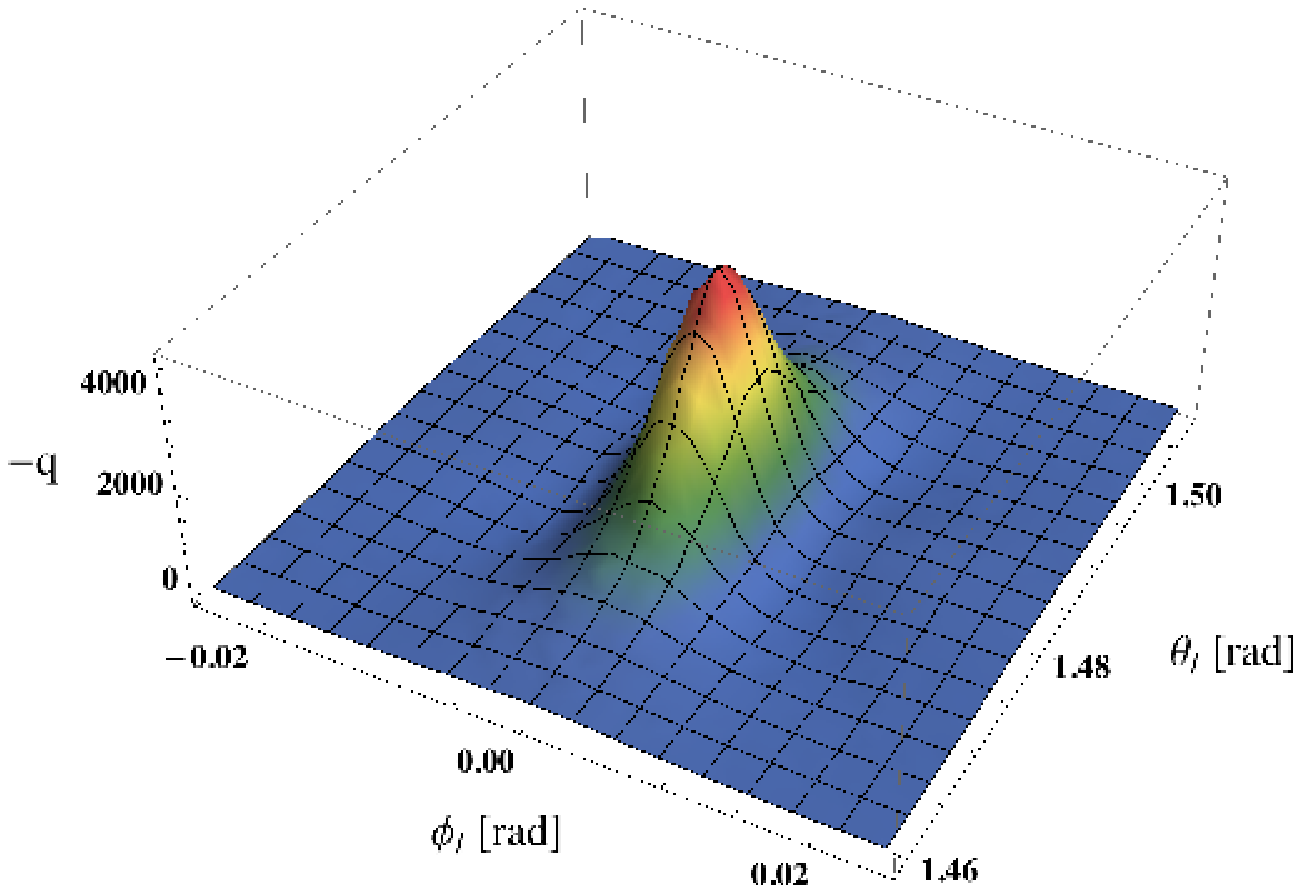}}
\end{minipage}
\hspace*{-1.6cm}
\begin{minipage}{8cm}
\resizebox{12cm}{!}{\plotone{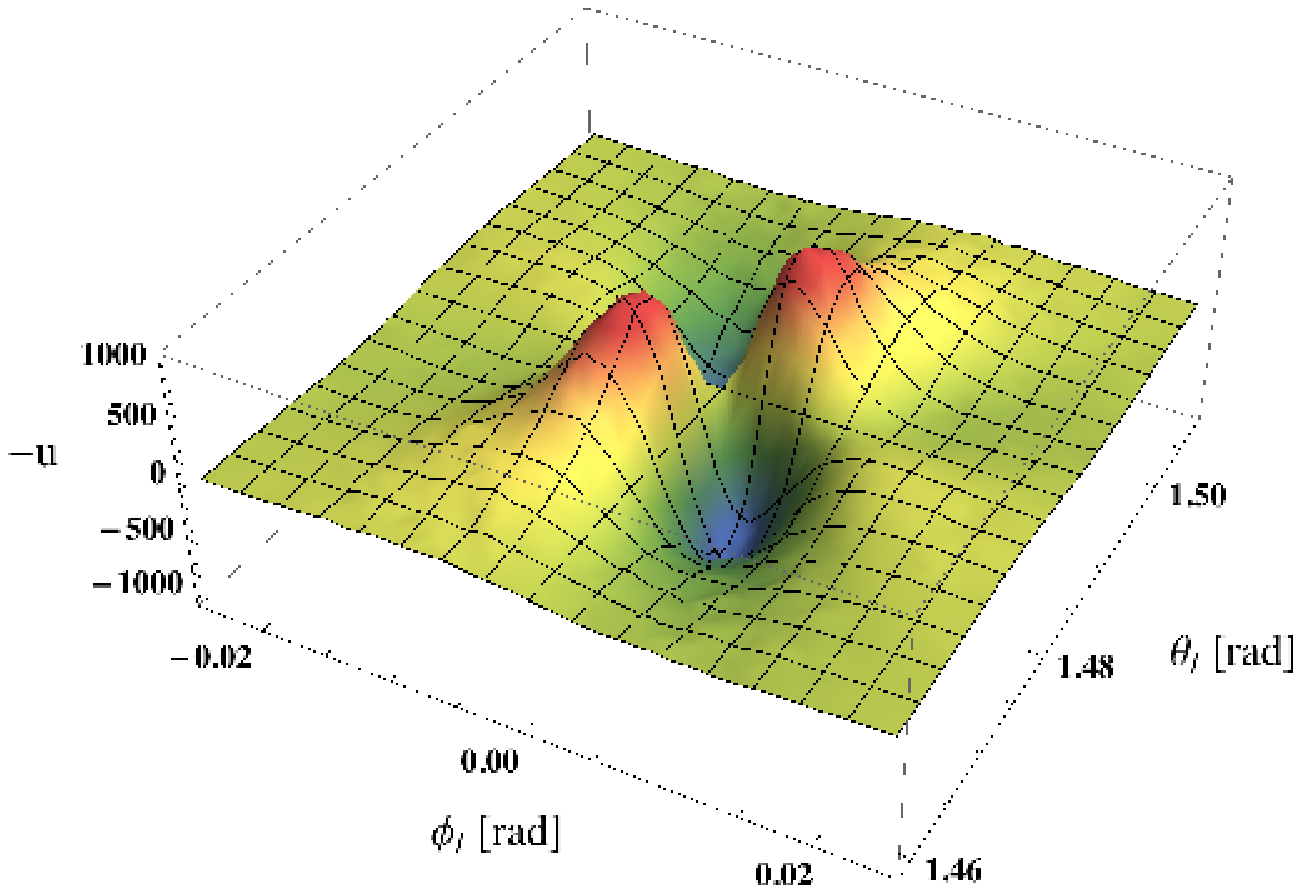}}
\end{minipage}
\begin{minipage}{8cm}
\hspace*{1.6cm}
\resizebox{10cm}{!}{\plotone{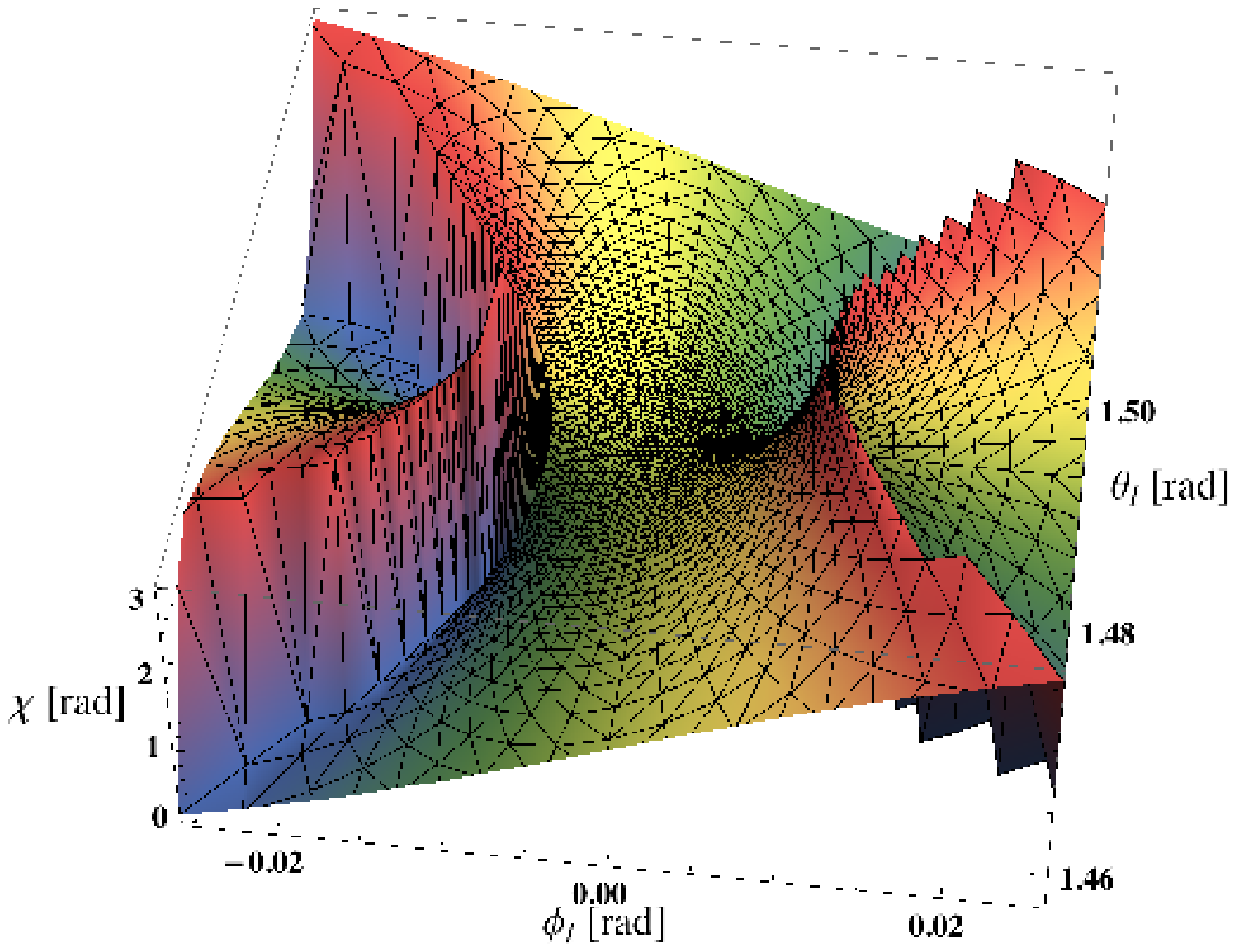}}
\end{minipage}
\caption{ \label{F:TP01} Distribution of the Stokes $i$, $q$ and $u$ 
parameters and the polarization direction $\chi$  of the inverse Compton 
emission from a monoenergetic unidirectional electron beam 
($\gamma=100$, $(\theta_{\rm e},\phi_{\rm e})=$ $(85^{\circ},0)$) 
scattering a monoenergetic unidirectional target photon beam 
propagating along the $z$-axis and polarized along the $y$-axis
($\omega=10^{12}$~Hz, $\mh{k}=(0,0,1)$, ${\bf s}=(1,1,0)$)
as function of the polar coordinates of the scattered
photon direction $\mh{l}$. $\chi$ is measured relative to $\mh{q}_+$
aligned with the projection of $\mh{k}$ onto the sky. 
}
\end{figure}

It is instructive to look at the numerical results for a monoenergetic unidirectional photon
beam scattering off a monoenergetic unidirectional electron beam.
Figure \ref{F:TP01} shows the Stokes parameters $i$, $q$, and $u$ and the 
polarization direction $\chi$ for an electron beam with $\theta_{\rm e}=85^{\circ}$, 
$\phi_{\rm e}=0^{\circ}$ and $\gamma=100$ impinging on a photon beam 
with frequency $\omega=10^{12}$~Hz incident along the $z$-axis ($\mh{k}=\mh{z}$), 
100\% polarized along the $y$-axis with an initial Stokes vector ${\bf s}=(1,1,0)$. 
As explained in the previous 
section, the polarization direction is measured clockwise from the direction of 
$\mh{k}$ in the sky when looking into the inverse Compton scattered beam.

In the electron rest frame, the intensity of the Thomson scattered photons resembles 
the torus-shaped intensity emitted by a Hertzian dipole. 
After back-transformation into the PF, 
the intensity of the inverse Compton emission shows two poles with zero intensity 
at an angular distance of $2/\gamma$ from each other bracketing the high-intensity 
radiation emitted in the direction of the electron direction. 
The $q$ and $u$ distributions look complicated, but encode simple information:
the inverse Compton emission from monoenergetic unidirectional electrons is
everywhere polarized to 100\%, and beyond the poles of the Hertzian dipole,
the polarization direction exhibits a smooth rotation with a periodicity of 
$\pi/2$ around the maximum-intensity direction.
The frequencies of the scattered photons (not shown here) are approximately evenly 
distributed between $\omega_{\rm min}$ and $\omega_{\rm max}$ 
($dN_{\gamma}/d\omega_{\rm s}\approx const$).

\begin{figure}
\hspace*{-2.5cm}
\begin{minipage}{11cm}
\resizebox{13cm}{!}{\plotone{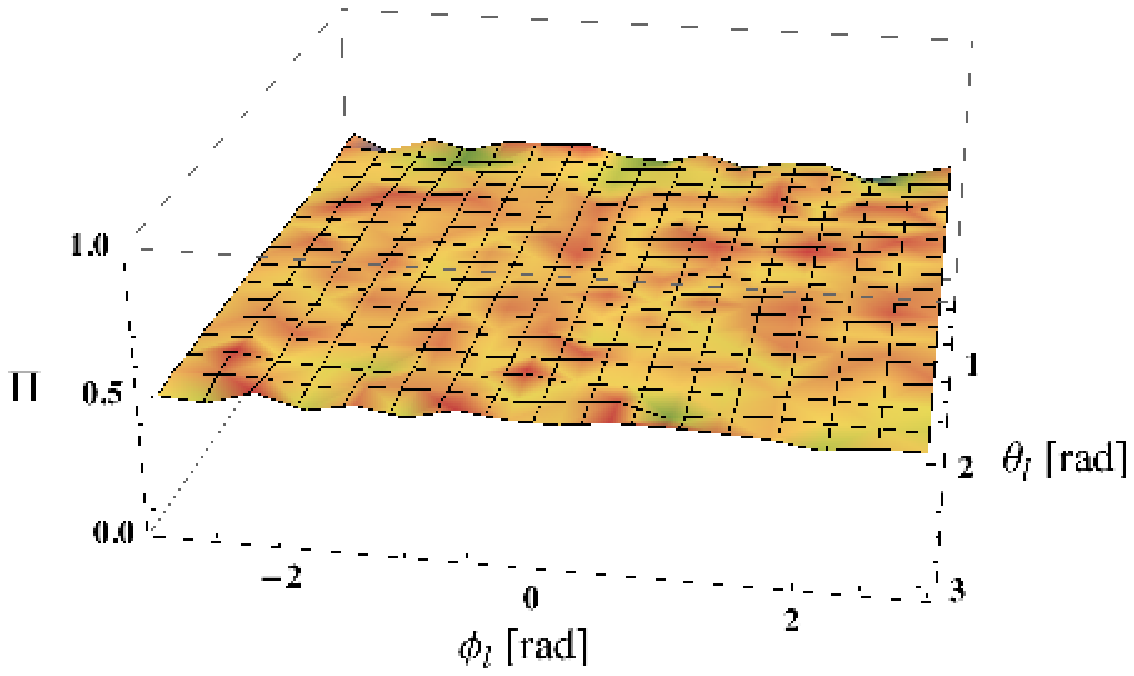}}
\end{minipage}
\begin{minipage}{8cm}
\hspace*{-1cm}
\resizebox{13cm}{!}{\plotone{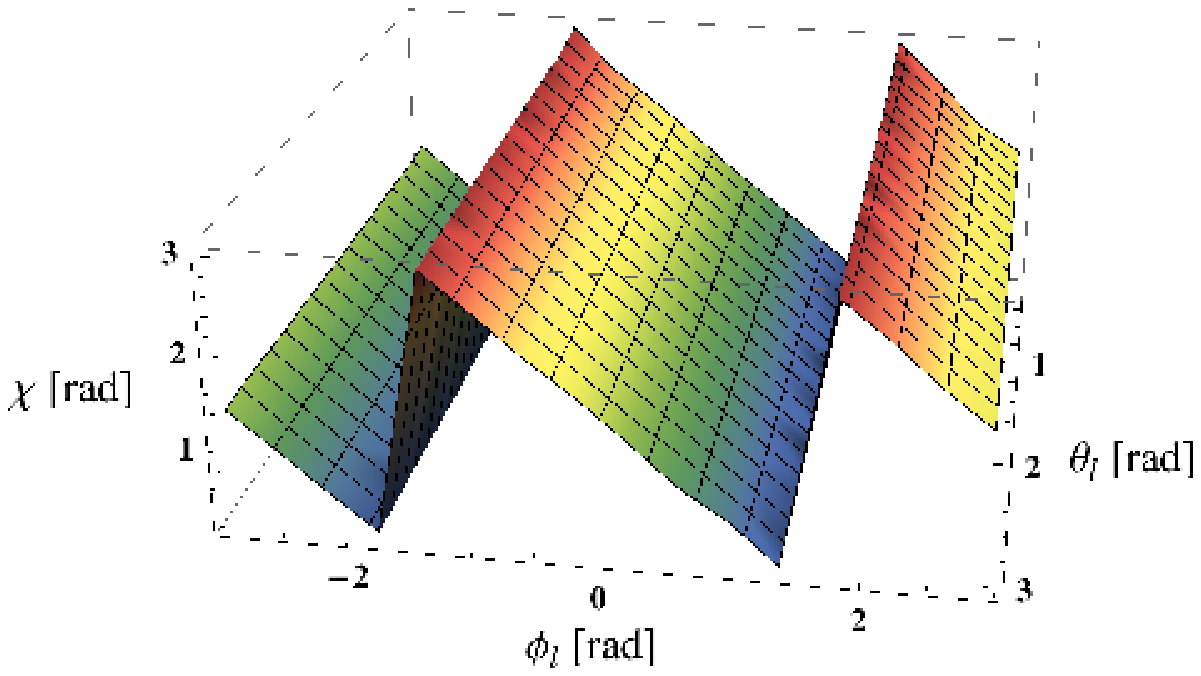}}
\end{minipage}
\caption{ \label{F:TP03} Polarization degree $\Pi$ and polarization
direction $\chi$ of the inverse Compton emission from
a monoenergetic isotropic electron distribution ($\gamma=100$)
scattering a monoenergetic unidirectional target photon beam 
($\omega=10^{12}$~Hz, $\mh{k}=$(0,0,1), ${\bf s}=(1,1,0)$)
as function of the polar coordinates of the scattered
photon direction $\mh{l}$. 
$\chi$ is measured relative to $\mh{q}_+$
aligned with the projection of $\mh{k}$ onto the sky. 
}
\end{figure}

For an isotropic electron configuration, electrons with initial directions 
within a few $\gamma^{-1}$ from the observed direction contribute $i$, $q$, and 
$u$-patterns as shown in Figure~\ref{F:TP01}. It is clear that summing up 
positive $i$-values and positive and negative $q$ and $u$-values will result 
in an overall polarization degree smaller than unity. Figure \ref{F:TP03} shows 
the intensity, polarization degree and polarization direction for an 
isotropic electron distribution with all other parameters being the 
same as in the previous example. 
The angular distribution of the inverse Compton emission (not shown here) is 
constant up to a dependence $\propto (1-\beta \mu_{\rm kl})$ 
(with $\mu_{\rm kl}=\mh{k}\cdot\mh{l}$) owing to the relative 
motion of the electrons and photons.
Figure \ref{F:TP03} shows the polarization degree and polarization direction
for all scattering directions. The simulations show that the polarization degree 
is $\Pi=0.5$ independent of the scattering direction and that the polarization 
direction depends on $\mh{k}$, $\mh{\boldsymbol\varepsilon}$, and $\mh{l}$ as predicted
by \eq{E:pd2}. 

\begin{figure}
\hspace*{-2.5cm}
\begin{minipage}{11cm}
\resizebox{13cm}{!}{\plotone{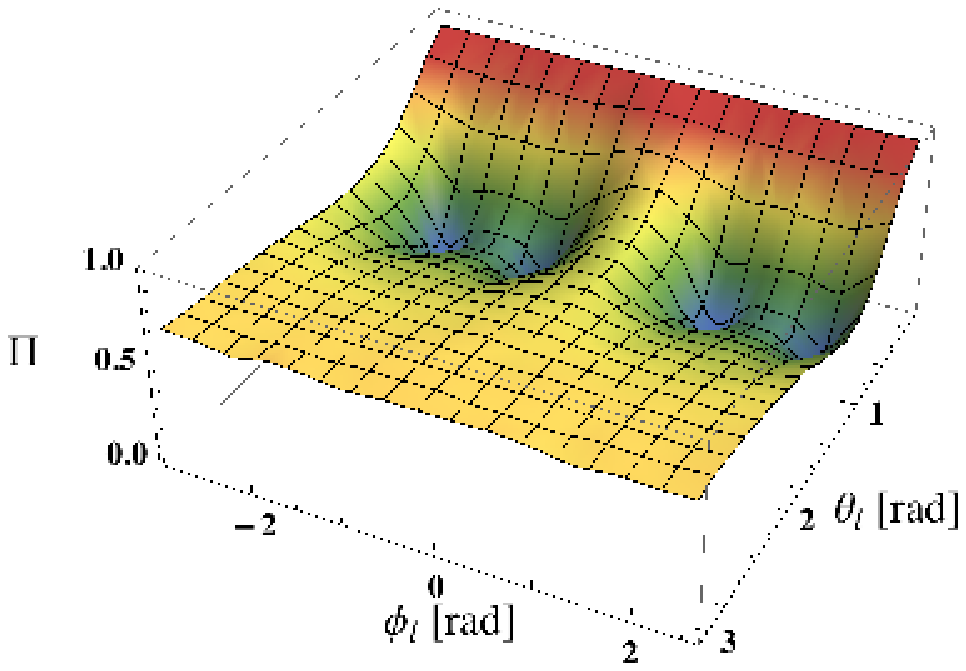}}
\end{minipage}
\begin{minipage}{8cm}
\hspace*{-1cm}
\resizebox{13cm}{!}{\plotone{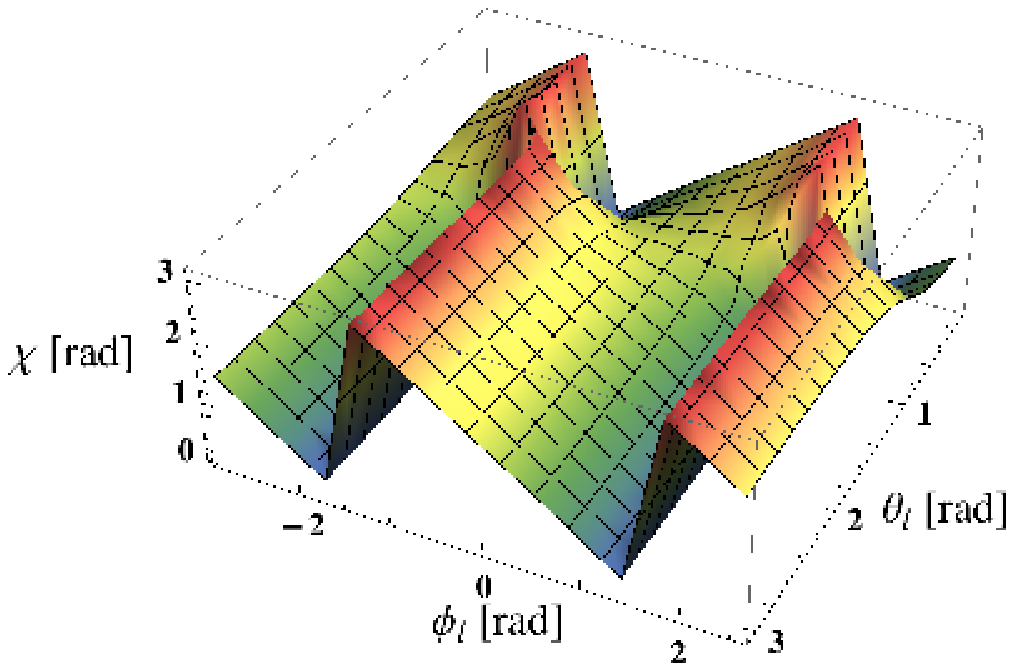}}
\end{minipage}
\caption{ \label{F:TP03b} Same as Figure \ref{F:TP03}, but for electrons with
Lorentz factor $\gamma=2$.}
\end{figure}

For mildly relativistic electrons, the polarization properties
do depend on the scattering direction. As an example, Figure \ref{F:TP03b} shows 
the same as the previous figure, but for $\gamma=2$.
At an angle $\theta_{\rm l}\approx$ 60\deg to the target photon beam,
the polarization degree dips and the dependence of $\chi$ on 
the azimuthal angle of $\mh{l}$ reverses.
The main difference between the mildly relativistic and highly relativistic cases is 
that in the latter $\mh{p}$ and $\mh{k}'$ are almost always nearly antiparallel
while in the former, the photons approach the electrons from a wide range of directions.
In the mildly relativistic case, the polarization direction of the incident photon is 
thus not ``rotated'' around an axis perpendicular to the $\mh{p}$-$\mh{l}$ plane. 

\begin{figure}
\begin{minipage}{8cm}
\resizebox{8cm}{!}{\plotone{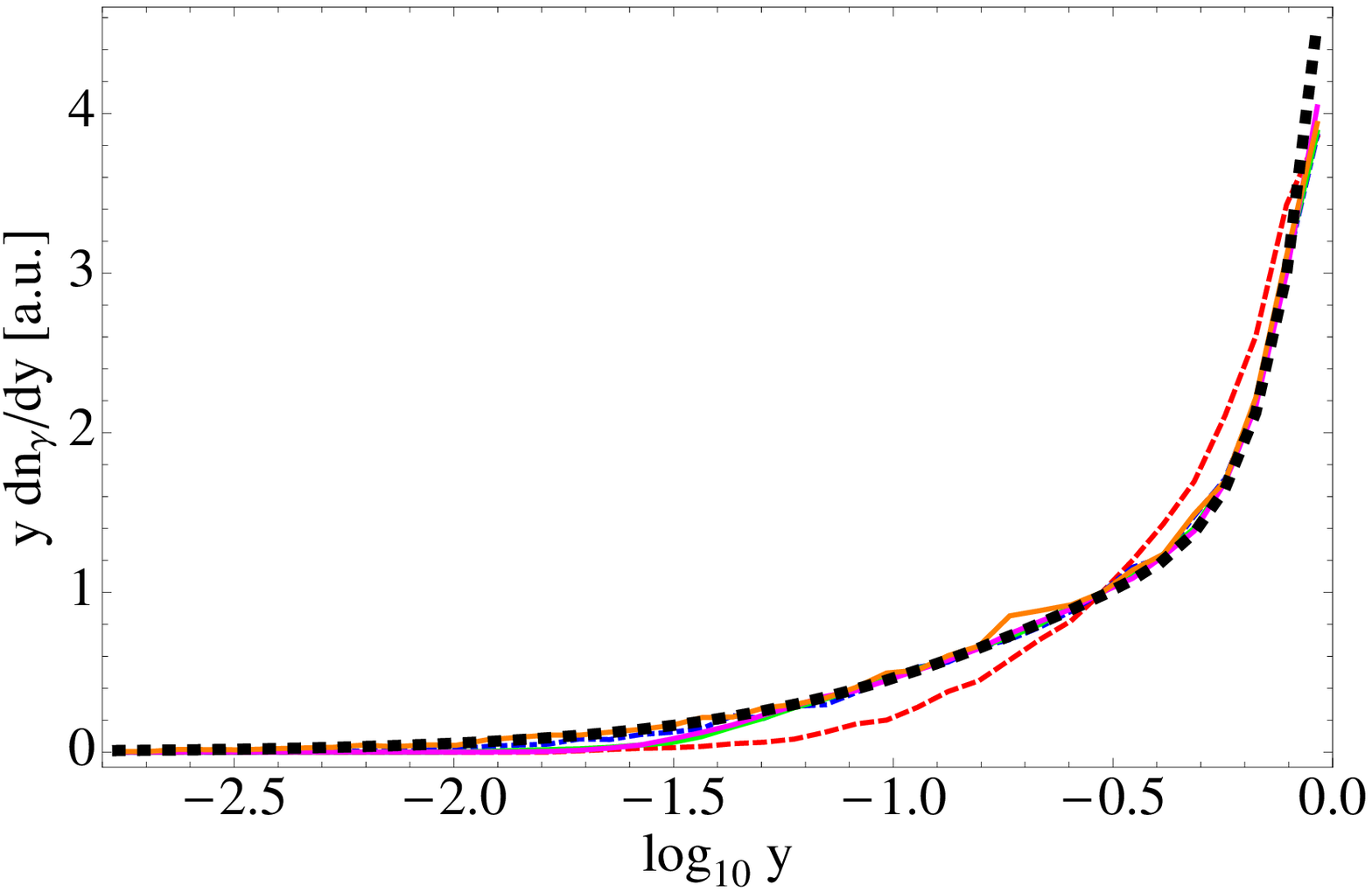}}
\end{minipage}
\begin{minipage}{8cm}
\resizebox{8cm}{!}{\plotone{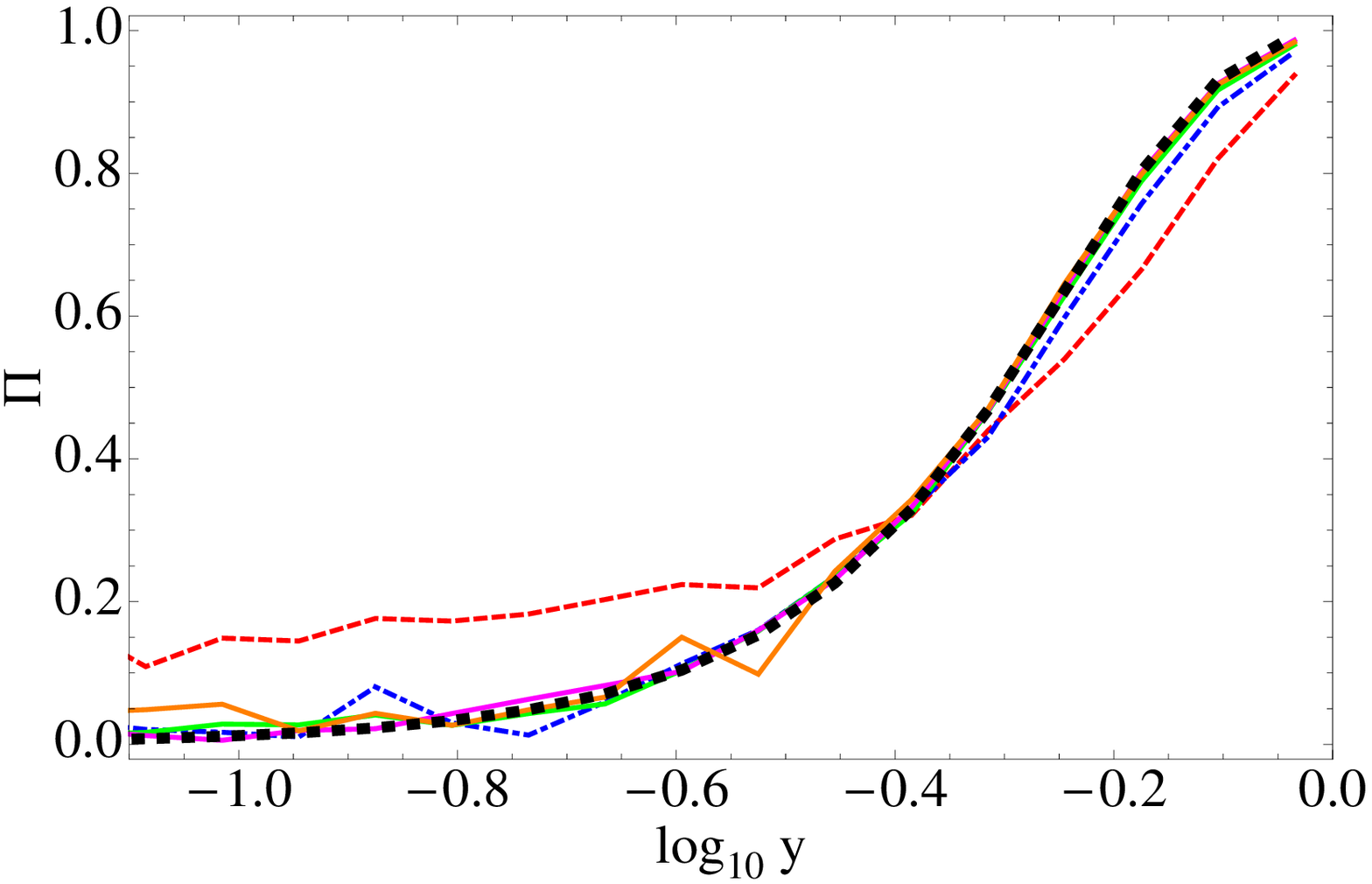}}
\end{minipage}
\caption{ \label{F:TP05b}
Frequency dependence of the intensity (left panel) and 
polarization degree (right panel) of the 
inverse Compton emission from a monoenergetic isotropic 
electron population scattering a monoenergetic unidirectional
photon beam. All parameter values are the same as in 
Figure \ref{F:TP01}, except $\gamma=2$ (dashed red lines),
$\gamma=5$ (dot-dashed blue lines),
$\gamma=10$ (solid green lines), 
$\gamma=20$ (solid magenta lines), and
$\gamma=100$ (solid orange lines).
The dotted black lines display the analytical results from \eqs{E:in}{E:pi}.
The frequencies are given as $y=\omega_{\rm s}/\omega_{\rm max}$ 
in units of the maximum frequency for Thomson processes. 
The results are shown for photons emitting along the 
direction $(\theta_{\rm l},\phi_{\rm l})=$ $(85^{\circ},0)$.
The plotting range of the right panel is restricted to the range
where the statistical errors on the polarization degree are
sufficiently small (a color version of this figure is available in the online journal).
}
\end{figure}

In the next step, we compare the frequency dependence of the intensity and polarization degree 
and direction with the theoretical predictions from Equations (\ref{E:in}), (\ref{E:pi}), 
and (\ref{E:pd1}-\ref{E:pd2}), respectively. We consider the radiation scattered into the direction $\mh{l}$ 
with the polar coordinates $(85^{\circ},0^{\circ})$ for $\gamma$-values of 2, 5, 10, 20, and 100.
Figure \ref{F:TP05b} shows that \eq{E:in} describes the frequency dependence  
extremely well except for mildly relativistic electrons with $\gamma=2$. 
Furthermore, \eq{E:pi} gives a good description of the polarization degree -- again, 
except for $\gamma=2$.
Additional simulations at higher $\gamma$-values 
($\gamma=$ 2,500, 12,500, and 62,500) show intensity and 
polarization degree distributions identical to the one shown for
$\gamma=$100. We also verified that the polarization direction 
(not shown here) is independent of frequency.

The case of a powerlaw distribution of electrons ($dN_{\rm e}/d\gamma \propto \gamma^{-p}$
from $\gamma_1$ to $\gamma_2$) scattering off monoenergetic unidirectional photons is discussed 
in the paper of BCS. In the region where the emitted energy spectrum is a power-law, 
the polarization degree is predicted to be
\begin{equation}
\Pi_{\rm pl}\,\approx\,\frac{(1+p)(3+p)}{11+4p+p^2}.
\label{E:pl0}
\end{equation}
\eqs{E:in}{E:in} give a frequency dependent polarization degree of 
\begin{equation}
\Pi_{\rm pl}\,=\,
\frac{
\int_{\gamma_1{}'}^{\gamma_2}\,
\frac{dN_{\rm e}}{d\gamma}\,
\rho_{\gamma_0}\,
\Pi_{\gamma_0}\,
d\gamma_0}
{\int_{\gamma_1{}'}^{\gamma_2}\,
\frac{dN_{\rm e}}{d\gamma}\,
\rho_{\gamma_0}
\,d\gamma_0}
\label{E:pl}
\end{equation}
with $\gamma_1{}'={\rm Max}(\gamma_1,\gamma_{\rm min})$.
The equation is valid for $\gamma_{\rm min}<\gamma_2$; 
no emission is found at frequencies with $\gamma_{\rm min}>\gamma_2$.
For $\gamma_1<\gamma_{\rm min}$ and $\gamma_2\rightarrow\infty$, 
we recover \eq{E:pl0}. 

\begin{figure}
\begin{minipage}{8cm}
\resizebox{8cm}{!}{\plotone{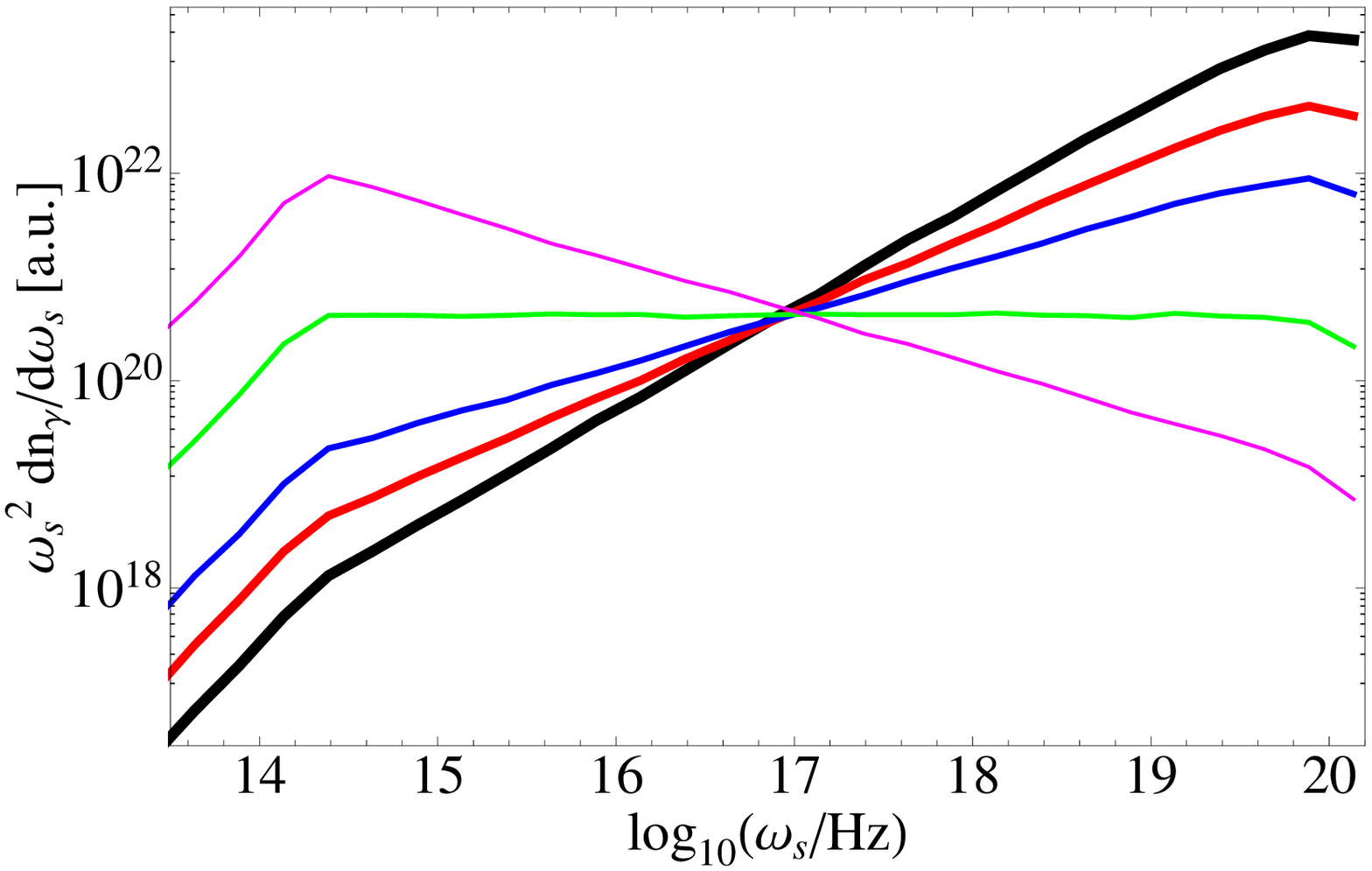}}
\end{minipage}
\begin{minipage}{8cm}
\resizebox{8cm}{!}{\plotone{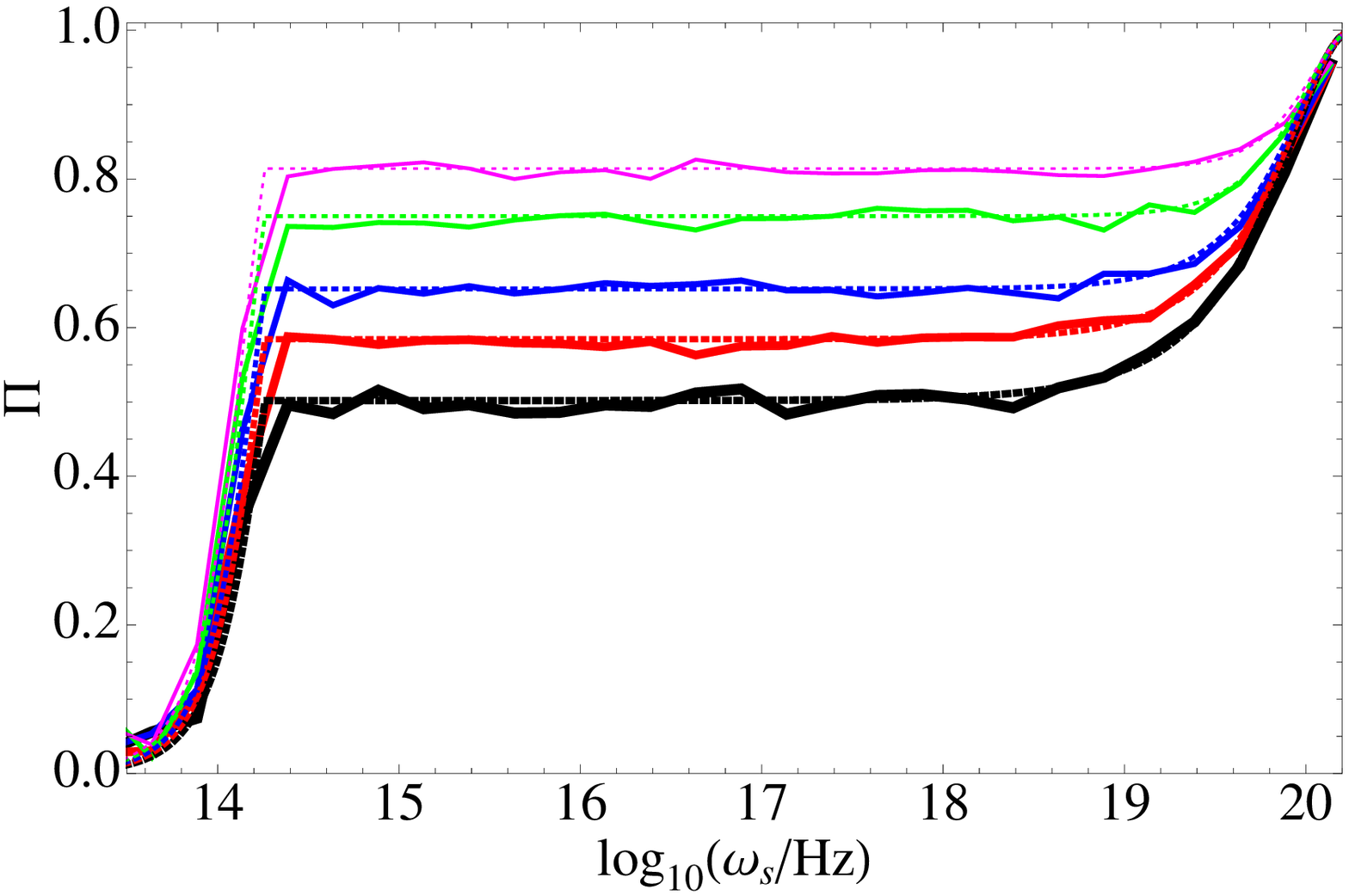}}
\end{minipage}
\caption{ \label{F:TP06} Intensity per logarithmic energy interval (left panel) and polarization 
degree (right panel) of the inverse Compton emission from isotropic power law electron populations
($dN_{\rm e}/d\gamma\propto \gamma^{-p}$ from $\gamma_1=10$ to $\gamma_2=10^4$) 
scattering a monoenergetic unidirectional target photon 
beam ($\omega=10^{12}$~Hz, $\mh{k}=$(0,0,1), {\bf s}=(1,1,0))
into the direction $(\theta_{\rm l},\phi_{\rm l})=$ $(85^{\circ},0)$.
The thick to thin solid lines display the simulation results for electron spectral 
indices of $p=$ 1.01, 1.5, 2, 3, 4. In the right panel, the dotted lines show 
the semi-analytical results from \eq{E:pl}.
}
\end{figure}

Figure \ref{F:TP06} compares the prediction from \eq{E:pl} with simulated results for
$\gamma_1=10$, $\gamma_2=10,000$, and electron indices $p$ between 1.01 and 4. 
The energy spectra of the inverse Compton emission indicate three regimes: a low-frequency regime with 
$\gamma_{\rm min}<\gamma_1$, a power law regime starting at 
$\gamma_{\rm min}$\uga$\gamma_1$, and a high-frequency regime with 
$\gamma_{\rm min}\rightarrow\gamma_2$.
In the low-frequency regime the polarization degree is low as the emission is produced
at low $y$-values. In the power-law regime, the polarization is independent of frequency
and agrees well with \eqs{E:pl0}{E:pl}. The softer the energy spectrum, the higher 
is the polarization degree. In the high-frequency regime, 
the polarization increases towards $\Pi=1$ as $\omega_{\rm s}\rightarrow \omega_{\rm max}(\gamma_2)$ 
because the emission is produced at $y$-values close to 1. The results validate the predictive 
power of \eq{E:pl0} for the power law regime, and of \eq{E:pl} for all frequencies.

We now turn our attention to the polarization of SSC emission from an isotropic electron population 
emerged in an uniform magnetic field {\bf B}. We assume that the electron and
synchrotron number densities are well described by power law distributions 
$dN_{\rm e}/d\gamma \propto \gamma^{-p}$ and $dN_{\gamma}/d\omega \propto \omega^{-(\alpha+1)}$
over the relevant energy and frequency ranges. The main objective of our discussion is to explain 
the polarization degree of the SSC emission from the polarization properties of the 
synchrotron emission and the general results concerning the depolarization effect 
of inverse Compton processes described above. 

As discussed in Section \ref{S:sc}, the intensity of the synchrotron emission is 
proportional to $\sin^{\alpha+1} \,\theta_{\rm B}$, and its polarization degree is $\Pi_{\rm S}$. 
In the power-law regime of the SSC radiation, the inverse Compton scattering 
reduces the polarization degree of synchrotron photons of a certain fixed frequency 
and propagation direction by the factor $\Pi_{\rm pl}$.
As the polarization direction of the inverse Compton emission depends on the
propagation direction and polarization direction of the synchrotron photons,
the net-polarization degree of the emerging SSC emission will be lower 
than $\Pi_{\rm pl}\Pi_{\rm S}$ and we write
\begin{equation}
\Pi_{\rm SSC}\,\approx \,\eta\,\Pi_{\rm pl}\,\Pi_{\rm S}.
\label{E:sc1}
\end{equation}
The factor $\eta$ depends on the angle $\vartheta_{\rm B}$ between the magnetic 
field and the line of sight, because the magnetic field direction determines the 
intensity and polarization pattern of the synchrotron emission. 
Furthermore, $\eta$ depends on the spectral index $\alpha$ of the synchrotron emission because
of Doppler boosting effects. 

We can derive an estimate for $\eta$ in the following way. 
Relative to the coordinate system $C_{\rm lB}$ with the $y$-axis parallel to the 
{\bf B}-field projected onto the plane of the sky and the $z$-axis pointing towards 
the observer, the $u$-parameter of the SSC radiation vanishes owing to the symmetry 
of the problem. The factor $\eta$ is then proportional to the absolute value of the
$q$-parameter and can be computed by integrating the $q$-parameter over the 
angular distribution of the synchrotron emission. 
Note that $\mh{p}$ is approximately parallel to $\mh{l}$ for most of the contributing
electrons. The expression for $\eta$ reads:
\begin{equation}
\eta\,=\,
\frac
{\left|\,
\int\,d\Omega_{\mh{k}}\,(1-\beta\mu_{\rm kp})^{\alpha+1}\,(\sin{\theta_{\rm B}})^{\alpha+1}\,\cos{2\,\chi}
\,\right|}
{\int\,d\Omega_{\mh{k}}\,(1-\beta\mu_{\rm kp})^{\alpha+1}\,(\sin{\theta_{\rm B}})^{\alpha+1}}.
\label{E:sc2}
\end{equation}
The integrals run over the directions $\mh{k}$ of the synchrotron photons. 
The first two factors of the integrands are weighting factors. 
One power of $(1-\beta\mu_{\rm kp})$ accounts for the 
dependence of the Compton scattering probability on the relative velocities of the 
photons and electrons. The additional power $\alpha$ stems from the fact that 
the synchrotron photons have an EF frequency $\omega'\propto$  
$(1-\beta\mu_{\rm kp})$ according to \eq{E:w1}, and  
the number of target photons available at a certain $\omega'$ in the EF scales proportional 
to $\frac{dN(\omega')}{d\omega'} = \frac{dN\left(\omega(\omega')\right)}{d\omega}\frac{d\omega}{d\omega'}$
$\propto$ $(1-\beta\mu_{\rm kp})^{\alpha}$. 
The factor $(\sin{\theta_{\rm B}})^{\alpha+1}$ accounts for the dependence
of the synchrotron emissivity on the angle $\theta_{\rm B}$ between $\mh{k}$ and {\bf B}. 
$\chi$ is the polarization 
direction of the inverse Compton scattered emission computed with the help 
of \eq{E:pd1}, and we made use of the identity $q=\cos{2\chi}$ from \eq{E:mchi}. 
The denominator normalizes the expression.
The numerical evaluation of the integral in the nominator gives negative values, 
corresponding to parallel synchrotron and SSC polarization directions --
perpendicular to the projection of the magnetic field in the PF onto the sky. 

A frequency dependent estimate of the polarization degree 
can be derived from \eqs{E:in}{E:pi}:
\begin{equation}
\Pi_{\rm SSC}(\omega_{\rm s})\,=\,
\frac{|q|}{i}\,=\,
\frac{\left|\,
\int\,d\Omega_{\mh{k}}\,
\int_{\omega_1}^{\omega_2}\,d\omega
\int_{\gamma'_1}^{\gamma_2}\,d\gamma_0\,
\frac{dN_{\rm e}}{d\gamma}\,
\frac{dN_{\gamma}}{d\omega}\,
(1-\beta \mu_{\rm kp})\,
(\sin{\theta_{\rm B}})^{\alpha+1}\,
\rho_{\gamma_0}\,\Pi_{\rm S}\,\Pi_{\gamma_0}\,
\cos{2\,\chi}\,\right|}
{
\int\,d\Omega_{\mh{k}}\,
\int_{\omega_1}^{\omega_2}\,d\omega
\int_{\gamma'_1}^{\gamma_2}\,d\gamma_0\,
\frac{dN_{\rm e}}{d\gamma}\,
\frac{dN_{\gamma}}{d\omega}
\,
(1-\beta \mu_{\rm kp})\,
(\sin{\theta_{\rm B}})^{\alpha+1}\,
\rho_{\gamma_0}}
\label{E:sc3}
\end{equation}
The integrals run over the directions of the synchrotron photons, the
synchrotron photon frequencies and the electron Lorentz factors. 
Only electrons with Lorentz factors $\gamma\ge$ $\gamma'_1\,=$ Max($\gamma_1,\gamma_{\rm min}$) contribute and we set
the value of the inner integral to zero if $\gamma'_1\,>\gamma_2$.
The first three factors of the integrand weight according to the
electron and photon number densities and the interaction probability,
the fourth factor is proportional to the synchrotron emissivity,
and the fifth factor gives the photon density of the inverse
Compton emission.
In the integral of the nominator we multiply all these weighting factors with
the fractional polarization of the synchrotron photons and the
inverse Compton emission, and multiply with the contribution of the considered
emission to the $q$-parameter at frequency $\omega_{\rm s}$.

The numerical evaluation of the SSC polarization follows the description in Section \ref{S:sc}. 
Figure \ref{F:TP07-1} presents the intensity and polarization degrees of the SSC emission for  
$\omega_1=10^{9}$~Hz, $\omega_2=10^{13}$~Hz, $\gamma_1=10$, $\gamma_2=10^5$, 
for ($\alpha=0.5,\,p=2$) and $(\alpha=1,\,p=3)$, and various $\vartheta_{\rm B}$-values. 
After low values at the lowest frequencies, the polarization degree
of the SSC emission $\Pi_{\rm SSC}$ reaches a rather stable plateau with 
a gradual increase before an eventual peak at the highest frequencies.
The polarization degrees calculated with \eq{E:sc3} (Figure \ref{F:TP07-1}, right panel, dotted lines) 
agree well with the results from the Monte Carlo simulations.
The right panel also shows the SSC 
polarization degrees presented in BS and CM. Whereas the polarization degrees of 
BS agrees well with our calculations, those of CM deviate by up to $\sim$15\%. 
We explain the disagreement with the fact that CM measure the polarization 
degrees at frequencies where electrons with $\gamma<10$ still play an important role.

Figure \ref{F:TP07-2} compares the $\vartheta_{\rm B}$-dependence of the emitted 
radiation with the predictions of \eqft{E:sc1}{E:sc3}.
The Monte Carlo simulations and \eq{E:sc3} give frequency dependent polarization degrees,
and we used here the polarization degrees averaged from 10$^{16}$ to 10$^{18}$ Hz.
The Monte Carlo simulations and \eqft{E:sc1}{E:sc3} give very consistent results. 
\citet[][and private communication]{Pout:94} suggested 
that $\eta$ in \eq{E:sc1} is given by $\eta\,=\,\sin^2 \vartheta_{\rm B}$. 
Our calculations show that the polarization degrees indeed scale proportional to 
$\sin^2 \vartheta_{\rm B}$, however even for $\vartheta_{\rm B}=\pi/2$ we find that $\eta<$1. 
The parameterization $\Pi_{\rm SSC}=\eta_0\,\sin^2 \vartheta_{\rm B}\,\Pi_{\rm pl}\,\Pi_{\rm S}$ 
with $\eta_0=$ 0.79 for ($\alpha=0.5,\,p=2$) and $\eta_0=$ 0.85 for $(\alpha=1,\,p=3)$
gives a good description of the results shown in Figure \ref{F:TP07-2}.
\subsection{The polarization degree of the inverse Compton emission from
unpolarized target photons}
\label{S:CMB}
One important implication of the analytical calculations of BCS and the numerical
results shown in Figure~\ref{F:TP03}, is that the polarization degree of inverse Compton 
emission off unpolarized radiation fields vanishes as long as the electron
Lorentz factors are $\uga 10$. The reason is that the scattering of two beams 
polarized perpendicular to each other will give two inverse Compton beams polarized 
perpendicular to each other, adding up to a beam with a vanishing net-polarization.
We tested this prediction explicitly for a case similar to the one discussed by
\citet{McNa:09}. We consider a jet of an AGN at redshift $z=2$, scattering unpolarized 
CMB photons into the X-ray band. The jet bulk Lorentz factor is $\Gamma_{\rm j}=5$ 
and the jet is oriented at an angle of $\vartheta_{\rm j}=80$\deg towards the line of sight. 
The observed inverse Compton radiation leaves the jet at angle of 80\deg in the AGN frame (AF), 
and an angle of 166\deg in the PF. 

We generate an isotropic distribution of CMB photons in the AF with a frequency of 
$(1+z)\,1.6\times10^{11}$~Hz $=4.8\times10^{11}$~Hz, and Lorentz transform the photon directions 
and photon frequencies into the PF. As the polarization degree is Lorentz invariant, 
the target photons are unpolarized in the PF, and we draw a random polarization 
direction in the PF. Subsequently, we simulate the Compton scattering taking into 
account the interaction probability as function of the electron and photon
velocity vectors, and accumulate the Stokes parameters of the radiation emitted 
into the direction of the observer. In the last step, we Lorentz transform the 
frequency of the emerging radiation into the AF, and redshift it into the observer frame. 

We simulated 10 million events with an electron Lorentz factor of $\gamma=10,000$. 
The choice of $\gamma$ produces inverse Compton emission with $F_{\nu}$ and $\nu F_{\nu}$-peaks 
in the 1-10 keV energy range. We calculate a net-polarization of the inverse Compton emission 
of 0.26\%. Note that the polarization degree is positive definitive, so that even an arbitrary 
large number of simulations will always produce positive polarization degrees. 
We divided the simulated data set in 100 sub-sets, and verified that the 
$q$ and $u$ parameters of the sub-sets exhibited mean values statistically 
consistent with 0, as expected for zero net-polarization. These results 
do not confirm polarization degrees of $\sim$20\% of inverse Compton scattered 
unpolarized photons as reported by \citet{McNa:09}.
\subsection{Inverse Compton emission in the Klein-Nishina regime}
\label{S:kn}
The Klein-Nishina regime starts when the EF energy of the target
photon is of the order or exceeds the electron rest mass energy: $\epsilon'\uga 1$.
In the deep Klein-Nishina regime ($\epsilon'\gg 1$), Equations 
(\ref{E:w1}), (\ref{E:w2}), and (\ref{E:w3}) imply that
head to head collisions of electrons and photons 
are likely to produce photons with an energy close to that of the incoming electrons. 
If we use the Fano Matrix with
a Stokes vector of (1,1,0), we see that the intensity of 
the emission scattered into the direction of the scattering electron
scales with $\epsilon'$ as 
$\left(\frac{\epsilon'_{\rm s}}{\epsilon'}\right)^2$ $(1+\epsilon'-\epsilon'_{\rm s})$
$\approx \epsilon'^{-1}$, and its fractional polarization scales 
as $(1+\epsilon'-\epsilon'_{\rm s})^{-1}$ $\approx \epsilon'^{-1}$.
The third term of the $(1,1)$-element of the Fano matrix introduces
unpolarized emission and an emission pattern
which depends in the EF mainly on the polar scattering 
angle and not on the azimuthal scattering angle.
   
\begin{figure}
\hspace*{-2cm}
\begin{minipage}{8cm}
\resizebox{14cm}{!}{\plotone{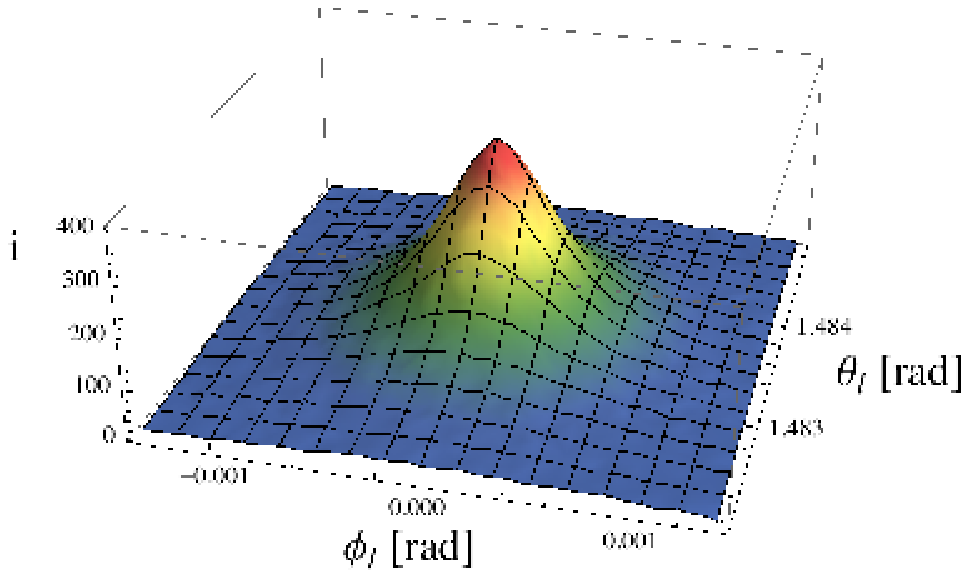}}
\end{minipage}
\begin{minipage}{8cm}
\hspace*{0.8cm}
\resizebox{14cm}{!}{\plotone{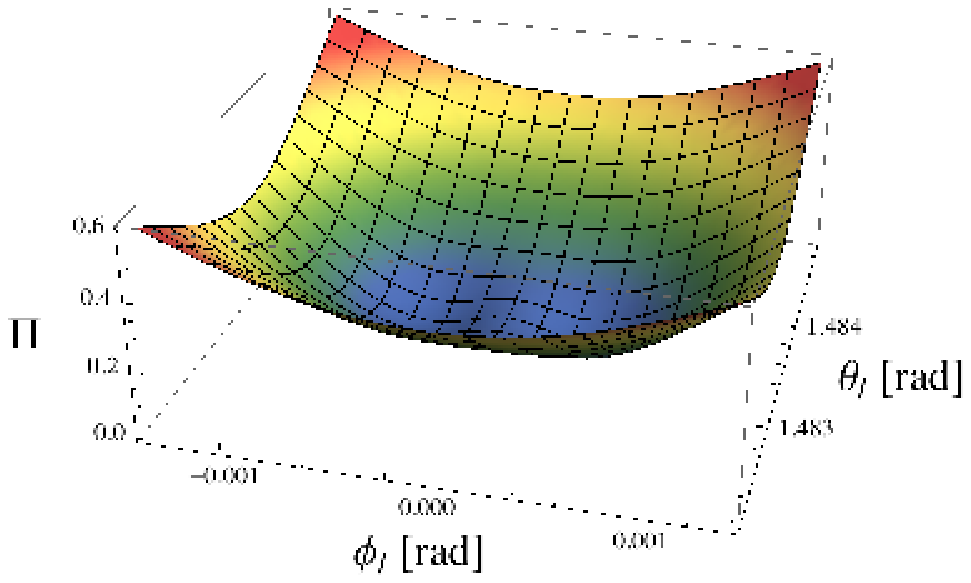}}
\end{minipage}
\caption{ \label{F:TP05e} Stokes $i$ parameter (left panel) 
and polarization degree $\Pi$ (right panel) of the inverse Compton
emission from a monoenergetic unidirectional electron beam
($\gamma=2,500$, $(\theta_{\rm e},\phi_{\rm e})=$ $(85^{\circ},0)$) 
scattering a monoenergetic unidirectional photon beam 
($\epsilon=0.01$, $\mh{k}=$(0,0,1), ${\bf s}=(1,1,0)$)
in the Klein-Nishina regime ($\epsilon'=$22.8).
}
\end{figure}

Figure \ref{F:TP05e} shows maps of the intensity distribution 
and the polarization degree of a monoenergetic unidirectional
photon beam with initial Stokes vector (1,1,0) scattered by 
monoenergetic unidirectional electrons.
We chose $\epsilon=1/100$ ($\sim$ 5.1 keV), 
$\mh{k}$ $=(0,0,1)$, $\gamma=2,500$ and 
$\mh{p}$ $=(\cos{85^{\circ}},0,\sin{85^{\circ}})$ 
which gives $\epsilon'=22.8$.
The intensity distribution is centrally peaked and decreases
rather monotonically with the distance from the peak.
The Hertzian dipole pattern is strongly suppressed.
As discussed above, the polarization degree at the peak of the 
intensity distribution is approximately 1/$\epsilon'$.
The polarization degree shows two dips with vanishing  
polarization degrees. These dips correspond to scattering directions
along the polarization direction of the target photon in the EF.
    
\begin{figure}
\resizebox{8cm}{!}{\plotone{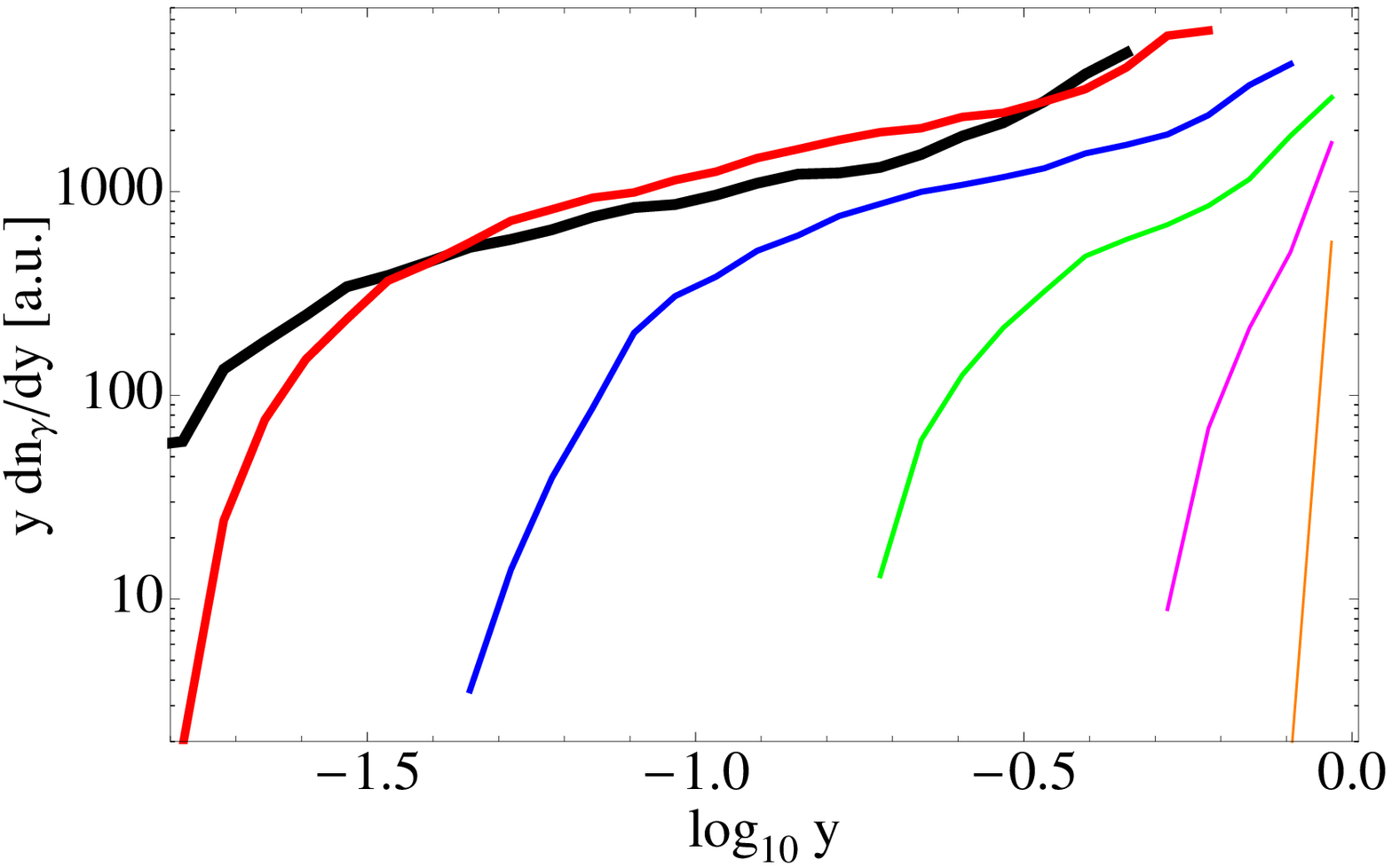}}
\resizebox{8cm}{!}{\plotone{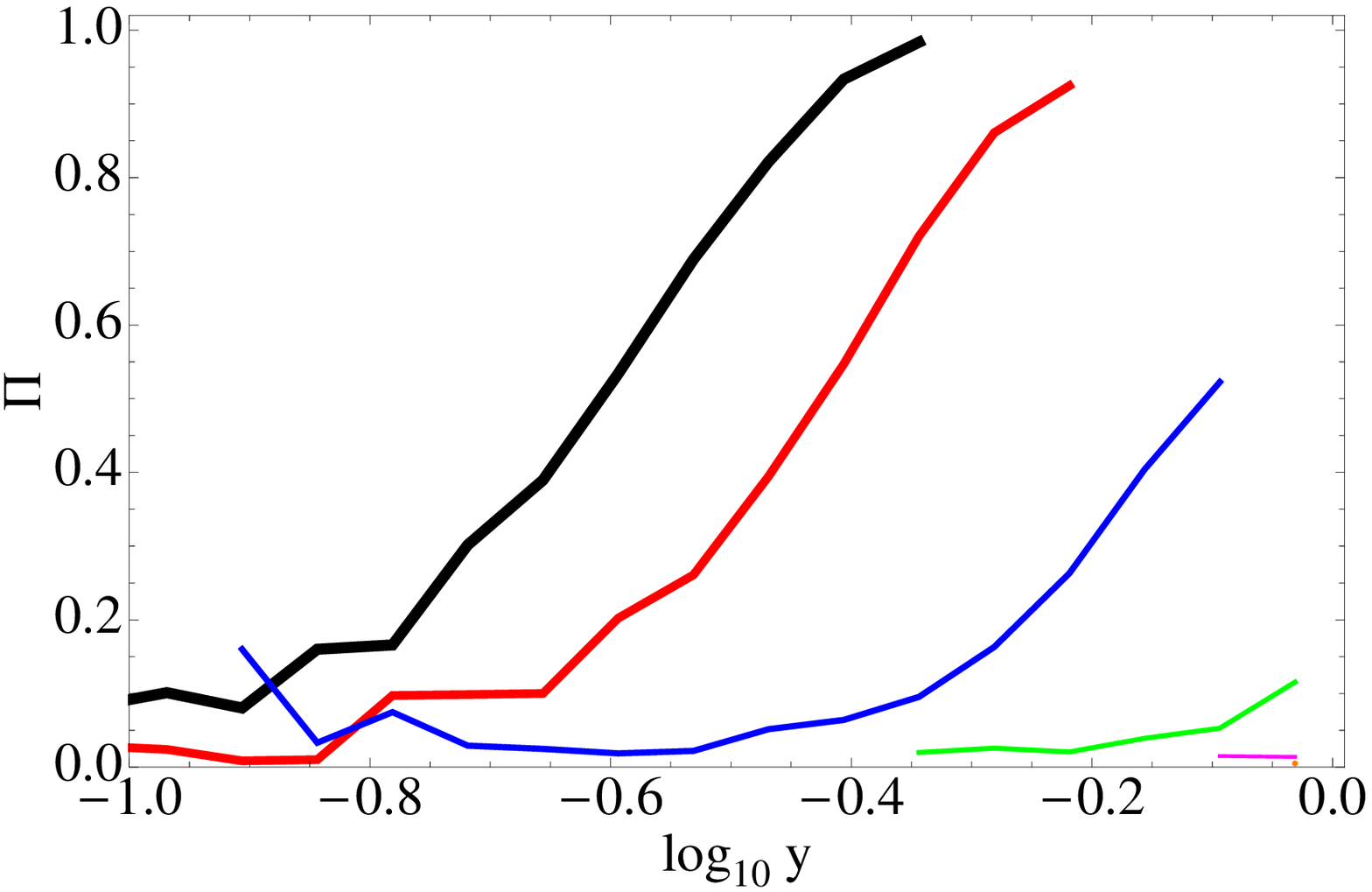}}
\caption{\label{F:TP05c} Intensity (left panel) and polarization degree (right panel)
of the inverse Compton emission in the direction
$(\theta_{\rm l},\phi_{\rm l})=$ $(85^{\circ},0)$ 
from monoenergetic isotropic electrons
scattering monoenergetic unidirectional photons
($\epsilon=0.0025$, $\mh{k}=$(0,0,1), ${\bf s}=(1,1,0)$).
From left to right, the lines display the results for
$\gamma$-values of 10, 100, 500, 2,500, 12,500, and 62,500.
In the EF, the target photon energy in units of the electron mass
is 0.02, 0.2, 1.1, 5.7, 29, 143 for the six simulated cases.
In the right panel, the frequencies  
are given in units of the maximum kinematically allowed frequency $y=\omega_{\rm s}/\omega_{\rm max,KN}$,
and the plotting ranges are restricted to the ranges
where the statistical errors on the polarization degree are
sufficiently small. 
}
\end{figure}

\begin{figure}
\resizebox{8cm}{!}{\plotone{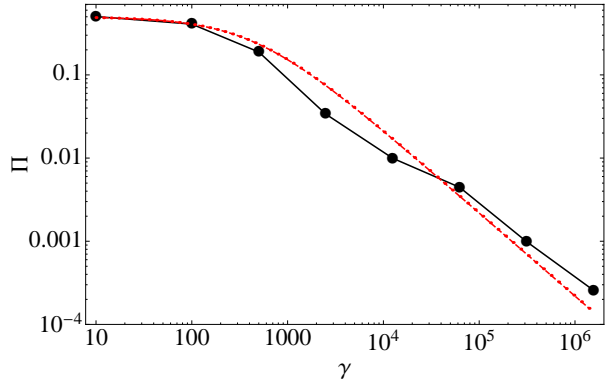}}
\caption{\label{F:TP05cc} 
Net-polarization degree of all emitted photons as function
of the electron Lorentz factor (solid line and circles) for the 
simulations of Figure \ref{F:TP05c} with additional simulations for
$\gamma$-values of $3.125 \times 10^5$ and $1.5625 \times 10^6$.
The dashed line displays the function $\Pi=0.5/(1+\epsilon')$ showing
that the net-polarization degree goes down approximately
as the inverse of the EF target photon energy $\epsilon'$. 
}
\end{figure}

Figure \ref{F:TP05c} presents the intensity and polarization
degree for a monoenergetic and unidirectional photon beam with an
initial Stokes vector (1,1,0) scattering off monoenergetic
isotropic electrons. We choose $\epsilon=1/400$ ($\sim$ 1.3 keV),
$\mh{k}$ $=(0,0,1)$, and an observer at $\mh{l}$ $=(\cos{85^{\circ}},0,\sin{85^{\circ}})$.
We simulate $\gamma$-values between 10 and 62,500 leading to
$\epsilon'$-values (computed with $\mu_{\rm kl}$ as proxi for $\mu_{\rm kp}$) 
between 0.02 and 143. All distributions are shown as function of the 
frequency $\omega_{\rm s}$ in units of the maximum frequency allowed kinematically
\begin{equation}
\omega_{\rm max,KN}=\frac{4 \gamma^2 \omega}{1+4\gamma\omega}.
\end{equation}
Deeper into the Klein-Nishina regime (at larger $\epsilon'$-values), the 
intensity distribution is more and more peaked towards $\omega_{\rm max,KN}$,
and the polarization is strongly suppressed.
Figure \ref{F:TP05cc} displays the net-polarization degree as a function
of $\gamma$. To guide the eye, the figure compares the numerical polarization degree
results with the function $\Pi=0.5/(1+\epsilon')$. The latter function agrees with the
numerical results for scattering in the Thomson regime, but deviates considerably
in the mild and deep Klein-Nishina regimes.
\section{Discussion: Application to GEMS observations}
\label{S:discussion}
In this paper, we describe a general formalism that can be used to study 
the polarization of inverse Compton emission, including SSC emission.
The comparison of the numerical results with the analytical results of BCS and BS
validate the analytical results for $\gamma$\uga~10. 
The numerical simulations show that Equations~(\ref{E:in}), (\ref{E:pi}), (\ref{E:sc3}),
and (\ref{E:pd1}-\ref{E:pd2}) can be used to predict the polarization properties of 
inverse Compton and SSC emission. An important qualitative result is that inverse 
Compton scattering of unpolarized target 
photons does not create a polarized signal as long as the electron distribution is 
isotropic over angular scales of $\gamma^{-1}$ and the electron Lorentz factors
exceed minimum values of $\sim$10. 
   
The GEMS X-ray polarimetry mission will have the opportunity to make major 
discoveries concerning AGNs, including constraints on the structure of
accretion disk coronae \citep{Schn:10}, the role and structure of the 
magnetic field in AGN jets, and the origin of the low-energy and high-energy 
components of the continuum emission from blazars \citep{Kraw:11}. 
Most relevant in the context of this paper are GEMS observations of the 
spatially unresolved continuum emission from blazar jets, because blazars are 
very bright X-ray sources. Two classes of objects are of particular interest.
Flat spectrum radio quasars (FSRQs) and low frequency and
intermediate frequency peaked BL Lac objects (LBLs and IBLs)
with energy spectra with a $\nu F_{\nu}$-peak in the IR/optical/UV band 
and one in the MeV to GeV band, and high frequency peaked BL Lac 
objects (HBLs) with one $\nu F_{\nu}$-peak in the UV/X-ray band and one
in the GeV/TeV band \citep[see][for compilations of spectral energy 
distributions]{Abdo:10a,Giom:02}. 

In the case of HBLs (e.g.\ the X-ray bright sources 
Mrk 421, Mrk 501, 1ES1959+650, PKS 2155-314, and PKS 1218+304), 
GEMS will sample the low-energy emission component, presumably of 
synchrotron origin. Optical blazar observations exhibit rather high polarization 
degrees, often close to the theoretical maximum $\Pi_{\rm S}$ for
synchrotron emission \citep[e.g.][]{Ange:80,Scar:97}. If X-ray emission from HBLs is also 
synchrotron emission, a commonly made assumption, GEMS observations
should reveal X-ray polarization degrees and directions similar
to those observed in the optical band. Indeed, the polarization 
degrees could be even higher, as X-ray emitting electrons radiatively 
loose their energy on shorter time scales than optically emitting
electrons. As the X-ray emitting electrons have thus less time to travel 
away from the acceleration sites, the X-ray bright regions 
are expected to be smaller than the optically bright regions. 
The more uniform magnetic field in the smaller X-ray bright regions 
should lead to a higher polarization degree of the synchrotron emission. 
The same reasoning suggests the possibility that continuous swings of the 
polarization direction as occasionally observed in the optical \citep[e.g.][]{Mars:08}, 
might be observed more often in the X-ray band. The polarization
swings have been interpreted to arise from the movement of 
a helical magnetic field which threads the jet through stationary 
shocks \citep{Mars:08}. 
Thus, GEMS has the potential to deliver observational evidence for
a helical magnetic field at the bases of jets, a prediction of 
magnetic models of jet formation, acceleration, and collimation 
\citep[e.g.][and references therein]{Spru:10}.

\begin{figure}
\begin{minipage}{8cm}
\resizebox{8cm}{!}{\plotone{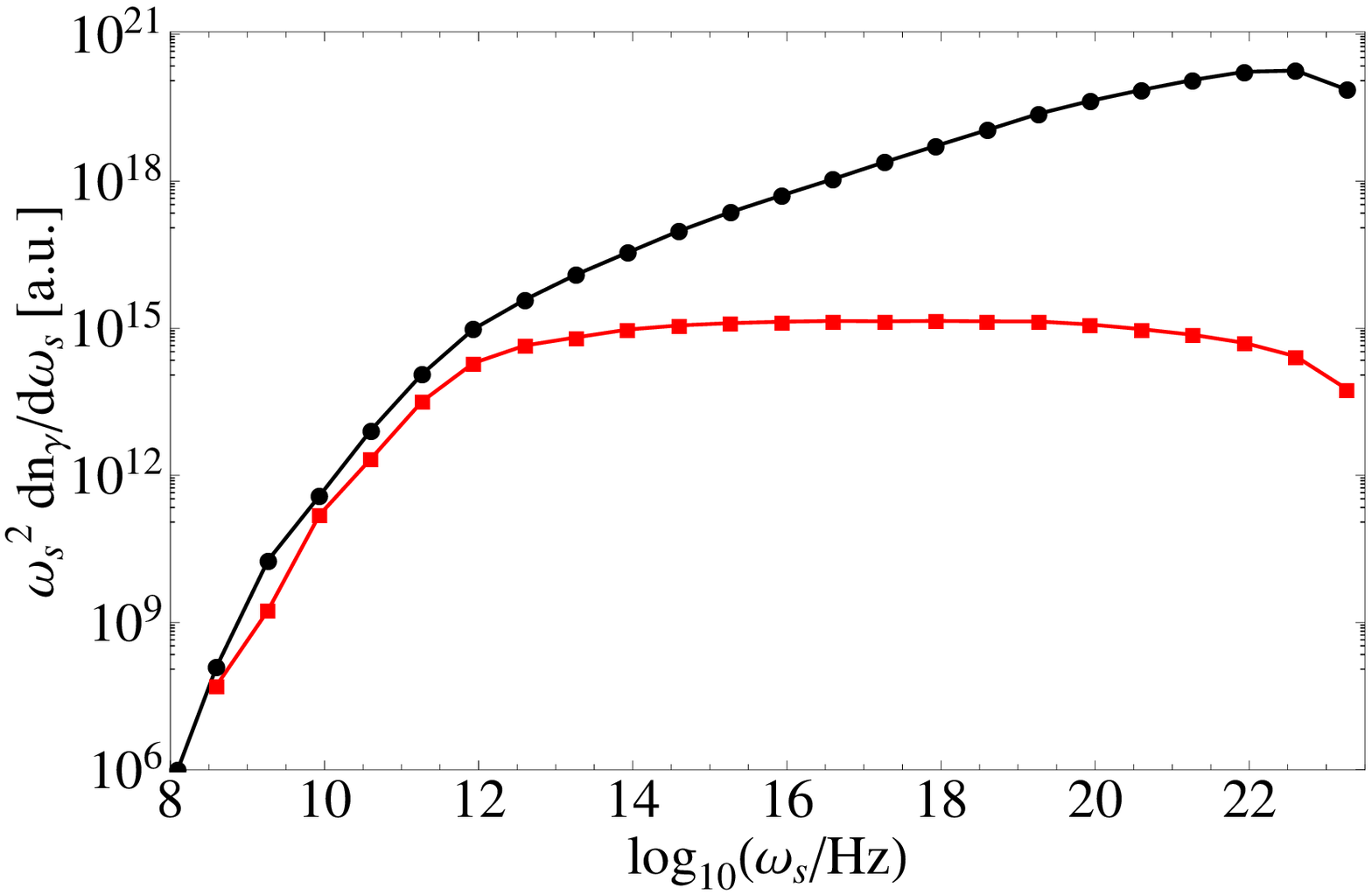}}
\end{minipage}
\begin{minipage}{8cm}
\resizebox{8cm}{!}{\plotone{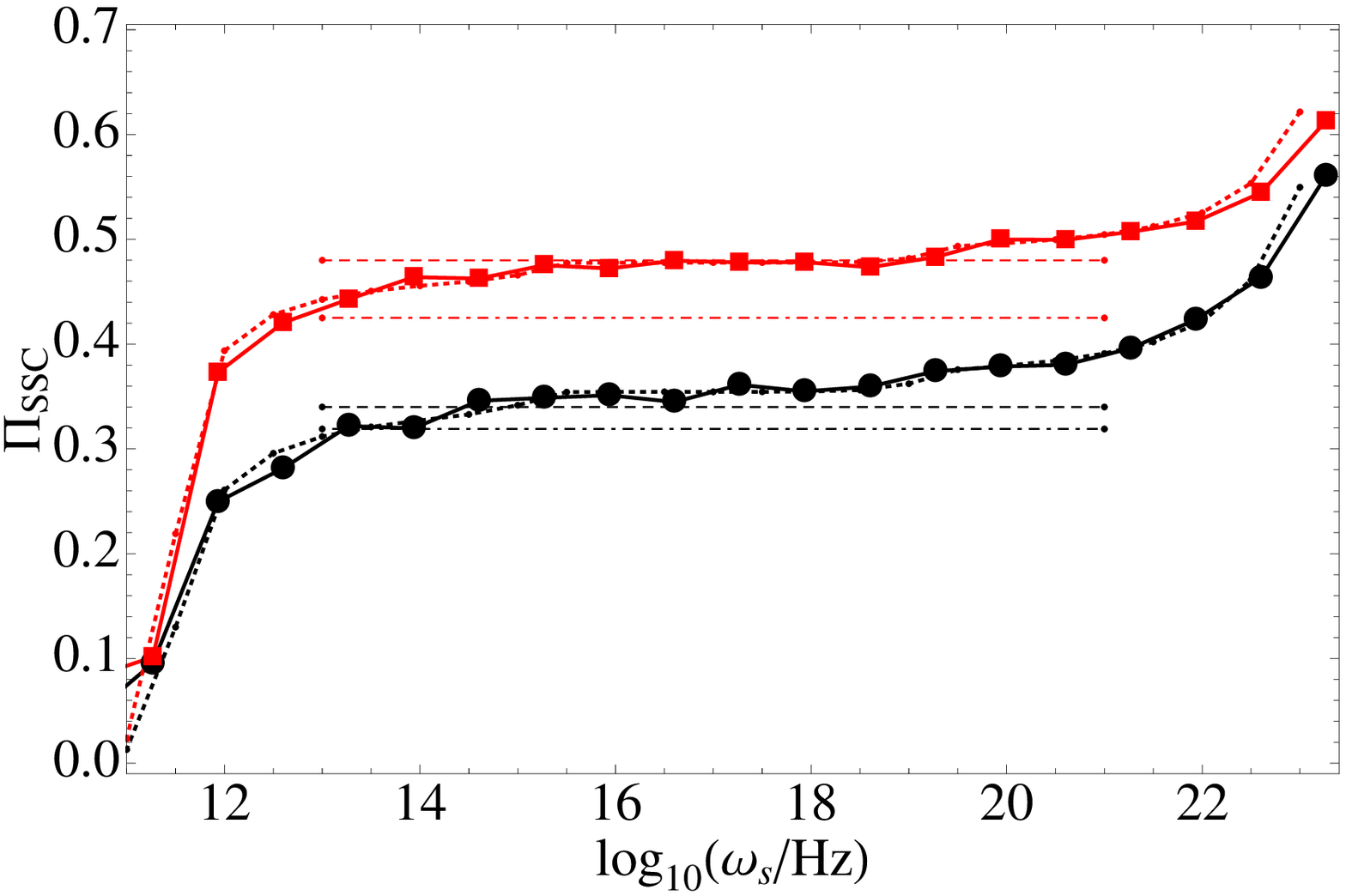}}
\end{minipage}
\caption{ \label{F:TP07-1}
Intensity per logarithmic energy interval (left panel) and polarization 
degree (right panel) of the SSC emission from isotropic power law electron populations
($dN_{\rm e}/d\gamma\propto \gamma^{-p}$ from $\gamma_1=10$ to $\gamma_2=10^5$) 
scattering synchrotron emission ($dN_{\gamma}/d\omega\propto \omega^{-(\alpha+1)}$
from $\omega_1=10^{9}$~Hz to $\omega_2=10^{13}$~Hz) from a {\bf B}-field
at 85$^{\circ}$ to the line of sight.
Numerical results are reported for $\alpha=0.5$ and $p=2$ (circles), 
and $\alpha=1$ and $p=3$ (squares).
The dotted lines in the right panel present the semi-analytical results from \eq{E:sc3}. The
dashed lines are the semi-analytical results from BS and the dashed-dotted 
lines report the numerical results from CM. Both papers do not give 
frequency resolved results.}
\end{figure}

\begin{figure}
\begin{minipage}{8cm}
\resizebox{8cm}{!}{\plotone{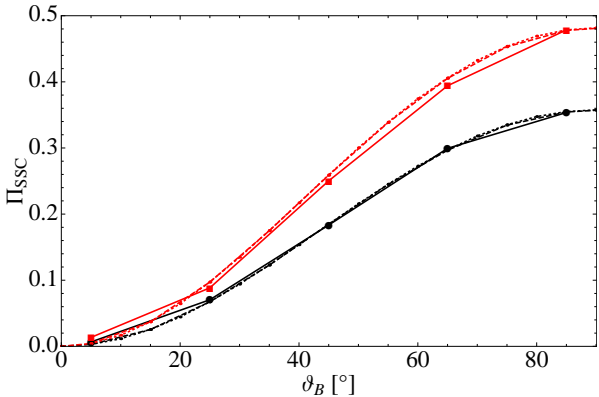}}
\end{minipage}
\caption{ \label{F:TP07-2}
Polarization degree of the SSC emission as function of the angle $\vartheta_{\rm B}$ 
between the magnetic field and the line of sight for $\alpha=1$ and $p=3$ (squares)
and $\alpha=0.5$ and $p=2$ (circles), evaluated between $10^{16}$ and $10^{18}$ Hz.
The dashed line displays the expectation from \eq{E:sc3}, and the dotted line shows the
results from \eqs{E:sc1}{E:sc2}.
}
\end{figure}

For FSRQs (e.g.\ the bright sources 3C~279, S~52116+81, and 1ES~0836+710), 
LBLs (e.g.\ the bright sources BL Lac, ON 231, and OJ 287), and some IBLs, 
GEMS will sample the high-energy $\nu F_{\nu}$-component, presumably 
of inverse Compton origin, and can be used to constrain the origin of 
this component. If the emission is of SSC origin, the inverse Compton emission
should track the polarization degree and polarization direction
of the synchrotron emission. Figures \ref{F:TP07-1} and \ref{F:TP07-2} show 
that the SSC emission can be highly polarized.
In external-Compton (EC) models, the dominant target photons
come from the accretion disk, from the corona, from the broad
line region clouds (BLR), or the CMB. As discussed in Section~\ref{S:thomson}, 
unpolarized target photons would result in very low ($\ll 1\%$)
polarization degrees. Although reflection off the BLR clouds
would polarize the target photons coming from a particular direction,
averaging over an axisymmetric cloud configuration would lead to
a low net-polarization of the target photons and thus to a rather low
polarization degree of the EC emission.
The detection of high polarization degrees would thus strongly 
favor a SSC origin of the X-ray emission. However, the detection
of vanishing or low polarization degrees would not exclude
the SSC scenario, as the SSC process can produce small polarization degrees
if the magnetic field is aligned with the line of sight.
The SSC hypothesis would even be consistent with low polarization degrees
observed for many sources, as it cannot be excluded that the blazar-zone
magnetic field is preferentially aligned with the line of sight.   
If the polarization degree is low but non-zero, simultaneous optical/X-ray 
measurements of the polarization degree and direction should be able
to decide between an SSC and an EC origin, as the former predicts
a good correlation while the latter does not.  

In the discussion so far, we have assumed that the magnetic field, and the
target photon and electron distributions do not change over the 
emission region. If information about the spatial (and/or temporal) 
distributions of these quantities is available, the polarization degree 
can be calculated with a calculation similar to \eq{E:sc3}.
In this case, one needs to integrate the expression in the nominator (inside the
absolute-value brackets) and in the denominator over the emission volume.
Inside the integrals, the magnetic field as well as the target photon and 
electron densities should be evaluated at the retarded times.
As \eq{E:sc3} gives only $|q|/i$, one has to calculate also $|u|/i$
by replacing $\cos 2\chi$ in the nominator by $\sin 2\chi$.
The polarization direction and degree can then be computed 
with \eqs{E:direction}{E:degree}, respectively. 
In most cases, a spatial variation of the source properties will lead to
a large number of free model parameters, and will render the unambiguous
interpretation of observational data more difficult.
Some of the most interesting GEMS results may come from observations
of polarization degrees close to the maximum theoretically possible
polarization degrees, which indicate a uniform emission region 
and make it possible to constrain models effectively.  
The reader interested in non-uniform emission regions should 
refer to the related discussions of the polarization of 
synchrotron emission from non-uniform emission regions 
\citep[e.g.][and references therein]{Bege:93,Rusz:02}.

The Monte Carlo approach presented in Section \ref{S:calculations} can also
be used to explore the polarization properties of inverse Compton
emission in the case that the electron phase space distribution 
is not isotropic in any frame of reference. 
Future research could combine the results of general relativistic or Newtonian 
magnetohydrodynamical simulations concerning the structure 
of blazar jets \citep[see e.g.][and references therein]{McKi:09,Meli:09}
with the results from particle-in-cell simulations of relativistic 
shocks \citep[see e.g.][and references therein]{Siro:09,Siro:11}
to derive estimates of the electron phase space distribution.
Monte Carlo simulations as described in this paper can then be used 
to derive observational signatures which can be compared to experimental data.
\acknowledgments
{\it Acknowledgments:}
HK thanks the members of the GEMS science working group, especially
M.\ Baring, J.\ Poutanen, and J. Scargle for discussions on the polarization degree of 
synchrotron and inverse Compton emission. The author acknowledges helpful
comments of an anonymous referee, and thanks A.~Parvulescu and J.~W.~Krawczynski 
for proofreading the manuscript. He acknowledges NASA for support from the 
APRA program under the grant NNX10AJ56G, the DOE for support from its 
high-energy physics division, and support from the McDonnell Center 
for the Space Sciences at Washington University.
 \\[2ex]

{\it Facilities:} \facility{GEMS}


%
%

\clearpage

\clearpage

\end{document}